\documentclass[twocolumn]{pasj00}

\usepackage{lscape}
\renewcommand{\tabcolsep}{4pt}

\begin{document}

\title{ Near infrared spectroscopy of M dwarfs. IV.\\ 
 A preliminary survey on the carbon isotopic ratio in
 M dwarfs
\thanks{Based on data collected at Subaru Telescope, which is operated 
by the National Astronomical Observatory of Japan.}\\
}

\author{Takashi \textsc{Tsuji}} %
\affil{Institute of Astronomy, School of Science, The University of Tokyo,
2-21-1 Osawa, Mitaka-shi, Tokyo, 181-0015}
\email{ttsuji@ioa.s.u-tokyo.ac.jp}


\maketitle

\begin{abstract}

Carbon isotopic ratios are estimated in 48 M dwarfs based on the medium 
resolution near infrared spectra ($ \lambda/\Delta\,\lambda \approx 
20000 $) of $^{13}$CO (3,1) band. We find clear evidence for the presence 
of a $^{13}$CO feature for the first time in the spectra of M dwarfs.  
Spectral resolution of our observed data, however, is not high enough to 
analyze the $^{13}$CO feature 
directly. Instead, we compare observed spectrum with  synthetic 
spectra assuming $^{12}$C/$^{13}$C = 10, 25, 50, 100, and 200 for each of
48 M dwarfs and estimate the best possible $^{12}$C/$^{13}$C ratio by
the chi-square analysis. The resulting $^{12}$C/$^{13}$C ratios in M dwarfs
distribute from 39 to a lower limit of 200. 
The mean value of 31 M dwarfs for which $^{12}$C/$^{13}$C ratios are 
determined (i.e., excluding those with the lower limit only) is 
$(^{12}{\rm C}/^{13}{\rm C})_{\rm dM} = 87 \pm 21$ (p.e.), and that of 
48 M dwarfs including those with the lower limit of 200 is 
$(^{12}{\rm C}/^{13}{\rm C})_{\rm dM} > 127 \pm 41$ (p.e.).
These results are somewhat larger than 
the $^{12}$C/$^{13}$C ratio of the present interstellar 
matter (ISM) determined from the  molecular lines observed in the millimeter  
and optical wavelength regions.  
Since the amount of  $^{13}$C in the ISM has increased with
time due to mass-loss from evolved stars, the $^{12}$C/$^{13}$C ratios 
in M dwarfs, reflecting those of the past ISM, should be larger than those of 
the present ISM.  In M dwarfs, log\,$^{13}$C/$^{12}$C 
plotted against log\,$A_{\rm C}$ shows a large scatter 
without clear dependence on the  metallicity. This result shows a marked 
contrast to log\,$^{16}$O/$^{12}$C (= log\,$A_{\rm O}/A_{\rm C}$)
plotted against log\,$A_{\rm C}$, which shows a rather tight correlation
with a larger value at the lower metallicity. 
Such a contrast can be
a natural consequence that $^{16}$O and $^{12}$C are the primary products 
in the stellar nuclear synthesis while $^{13}$C is the secondary product, 
at least partly.

\end{abstract}

\section{Introduction}

Besides the elemental abundances, isotopic ratios in stars and
interstellar matter (ISM) provide
important clues on  stellar evolution and  Galactic  chemical evolution.
However,  isotopic effects are prominent only in molecular
spectra but not in atomic spectra. For this reason, isotopic analysis has
mostly been done on molecular spectra observed in cool celestial objects
such as cool stars and the ISM. Because of this limitation, our knowledge on
the isotopic abundances in the Universe has  been rather poor  
compared to that on the elemental abundances. 
On the other hand, spectroscopic analyses of the isotopic ratios are 
somewhat simpler, compared to those of the elemental abundances,
in that the isotopic ratios are relatively insensitive to
the physical condition of the environments where spectral lines are formed.
For this reason, isotopic analyses have extensively been done on the spectra 
of cool stars  since the middle of the 20-th century even when analyses on
the elemental abundances were more difficult. 

For example, initial attempt to determine the $^{12}$C/$^{13}$C ratio
in cool giant stars has been done by \citet{Gre69}, who showed that the 
$^{12}$C/$^{13}$C ratio in red giant stars is decreased
compared to the solar system value of 89.9 \citep{And89}.  Then, extensive 
analyses on the $^{12}$C/$^{13}$C ratio in G and K giant stars have been 
done with the use of photoelectric scans of the CN red system (\cite{Day73}; 
\cite{Tom74}; \cite{Tom75}; \cite{Dea75}). The results confirmed
that the products of the CNO cycle are dredged-up in the red giant phase, but
the resulting $^{12}$C/$^{13}$C ratios appeared to be too small
compared with the theoretical predictions (e.g., \cite{Dea76}).
 Studies on  the $^{12}$C/$^{13}$C ratio were extended to red
supergiants \citep{Lam74} and to AGB stars  with the
use of the high resolution FTS spectra (\cite{Smi86}, \yearcite{Smi90};
\cite{Tsu08}). The results on oxygen-rich giants, supergiants, and AGB 
stars revealed  that these stars contribute to lower the $^{12}$C/$^{13}$C 
ratio in the ISM by their mass-loss.

Historically, it was on carbon stars that studies on carbon isotopes 
in stellar spectra have initiated: It was nearly a century ago when   
$^{13}$C bearing molecules were identified by the C$_2$
Swan bands, first as $^{12}$C$^{13}$C \citep{San29} and next as 
$^{13}$C$^{13}$C \citep{Men30} in carbon stars (also
referred to as R and N type stars). Initial works suggested that the
$^{12}$C/$^{13}$C ratios in most carbon stars are in the range of 2 -- 20, 
except for a few exceptional case (\cite{Mck60} and references cited 
therein). This result suggested that the 
$^{12}$C/$^{13}$C ratios in carbon stars may be related to the equilibrium 
value of the $^{12}$C/$^{13}$C ratio in the CN cycle. However, since the
$^{12}$C$^{12}$C Swan bands in carbon stars are very strong
and heavily saturated while the $^{12}$C$^{13}$C Swan bands less
saturated, the intensity ratio of $^{12}$C$^{12}$C/$^{12}$C$^{13}$C
should apparently be small leading to a low $^{12}$C/$^{13}$C ratio. 
For this reason,  the very low $^{12}$C/$^{13}$C ratios found in 
carbon stars can be due to imperfect 
correction for the saturation effect, and a possibility that
$^{12}$C/$^{13}$C $\approx 100$ was suggested for some cool carbon stars
(e.g., \cite{Fuj77}). Since then, many works on the  $^{12}$C/$^{13}$C ratio  
have been carried out for a large sample of carbon stars: \citet{Lam86} 
determined the $^{12}$C/$^{13}$C ratios in dozens of N type stars through
the analysis of the high resolution FTS spectra of the CN red system, CO
first and second overtones, and showed that  the  $^{12}$C/$^{13}$C ratios
in cool carbon stars extend to as large as the solar system value.
Determinations of the $^{12}$C/$^{13}$C ratios were extended through the
analysis of the CN red system to 62 N type stars by \citet{Ohn96} and 
to 44 carbon stars by \citet{Abi97}. The results on carbon-rich AGB 
stars revealed that the $^{12}$C/$^{13}$C ratios
in these stars are generally larger than in oxygen-rich AGB stars and
these stars contribute to increase the $^{12}$C/$^{13}$C ratio in the ISM 
by their mass-loss.

In contrast, isotopic analyses on cool dwarfs are quite limited, even
though extensive analyses on elemental abundances in late-type dwarfs   
have been done (e.g., \cite{Edv93}). It is true that molecular lines
are not so strong in F and G dwarfs. However, the major 
reasons for this contrast between high and low luminosity stars
may be that the high resolution spectroscopy
needed to isolate faint isotopic features could be applied more
easily to bright high luminous stars on one hand and also some isotopic 
features appeared to be enhanced in high luminous cool stars due to the
evolutionary effects in the stars themselves. 
 
In  M dwarfs, determinations of the
elemental abundances have also been  difficult, but we have shown
recently that the carbon and oxygen abundances in M dwarfs could be 
determined rather accurately by the use of the near infrared spectra of 
CO and H$_2$O, respectively (\cite{Tsu14}, \cite{Tsu15}, \cite{Tsu16}, 
hereafter Papers I, II, and III, respectively).
This has been made possible by several reasons: 
First, the well known difficulty of the continuum in cool stars
has been overcome in doing spectroscopic analysis by referring to the
pseudo-continuum, which can be evaluated accurately with the use of the recent
molecular line-list including many weak lines of H$_2$O (e.g., \cite{Bar06};
\cite{Rot10}). Then, spectroscopic analysis can be done essentially the
way as referring to the true-continuum, so long as the pseudo-continua can be
defined consistently in the observed and predicted spectra. Second, 
given that carbon and oxygen abundances are determined from stable molecules 
such as CO and H$_2$O, respectively, the problem of photospheric model should 
not be so serious. This is because CO consumes most of carbon and H$_2$O 
most of oxygen left after CO formation, and CO abundance is almost identical 
with that of C and H$_2$O abundance with that of O$-$C in M dwarfs.
For this reason, CO and H$_2$O abundances are insensitive
to  the uncertainties of the photospheric structures \footnote{
It is to be noted that this advantage does not apply to CO in general, but
applies to CO in  M dwarfs only. In warmer F or G dwarfs, CO
abundance is highly sensitive to temperature and cannot be used
for accurate abundance analysis (see footnote 9 in Paper I).}, and
carbon and oxygen abundances could be determined rather well despite
the use of model photospheres which anyhow cannot be very accurate
(as for further details, see e.g., Paper II).

Given that the elemental abundances of at least some elements
were well determined in M dwarfs, a further important possibility in
M dwarfs is that the isotopic ratios can also be determined from 
the molecular spectra, at the same
time as the elemental abundances in the same object. In the near
infrared spectra of M dwarfs, not only $^{12}$C$^{16}$O but also
$^{13}$C$^{16}$O bands are observed, and the $^{12}$C/$^{13}$C ratios 
can be discussed. A problem is that the spectral resolution of our spectra
is about $\lambda/\Delta\,\lambda \approx 20000 $ or the 
velocity resolution is about 16\,km\,s$^{-1}$. This resolution  
barely made it possible to measure equivalent widths and hence to
determine  the elemental abundances of carbon and oxygen
(Papers I, II, and III). However, this resolution is not
high enough to measure  line profiles accurately and hence to
analyze faint isotopic features. For this reason, our analysis
on the carbon isotopic ratios is necessarily preliminary.
Even with this limitation, however, we hope that 
we can  discuss the elemental and isotopic
abundances of carbon simultaneously in a large sample
of M dwarfs for the first time. 
 
In this paper, we first summarize the known data on our 
program stars from Papers I, II, and III in section 2. We then  
investigate  how to estimate  the $^{12}$C/$^{13}$C ratios
from the spectra of $^{13}$CO  in section 3.  
The resulting $^{12}$C/$^{13}$C ratios, with
 assessments of the accuracy, are given in section 4.
The resulting carbon isotopic ratios in M dwarfs are discussed in 
comparisons with those of the ISM in section 5.
 
\section{Preparatory data for isotopic analysis}

\subsection{Observed data}
  We use the observed data introduced in Paper I (see its Table\,1)
 and III (its Table\,1), in which we analyzed $^{12}$C$^{16}$O
 and H$_2$O lines. In this paper, we focus our attention to
$^{13}$C$^{16}$O features recorded on the 24-th order
(23437 -- 23997\,\AA) of the echelle
spectra observed with the Infrared Camera and Spectrograph 
(IRCS: \cite{Kob00}) of the Subaru Telescope. The resolution
$ \lambda/\Delta\,\lambda$ is about 20000 or the velocity resolution
is about 16\,km\,s$^{-1}$. In our observations, the targets were
nodded along the slit, and observations were taken in an ABBA sequence, 
where A and B stand for the first and second positions on the slit (Paper I).
For estimating the signal to noise (S/N) ratio,
the noise level is estimated from the difference of the spectra observed 
at A and B positions. The resulting S/N ratio is given in the last column 
of Table 3 for each object, and
the S/N ratios are mostly between about 50 and  150.

\subsection{Stellar data}

Our program stars consist of M dwarfs studied so far by us: We first
determined the fundamental parameters such as $T_{\rm eff}$ and
log\,$g$, prepared model photospheres, and
determined carbon abundances based on the $^{12}$C$^{16}$O lines for 
42 M dwarfs (Paper I). For 38 objects in these
42 M dwarfs, we also determined oxygen abundances from  H$_2$O lines
(Paper II). We then studied additional eight
late M dwarfs (Paper III). As a result,
we determined carbon and oxygen abundances in 38 + 8 = 46 M dwarfs. 

We did not determine oxygen abundances in four early M dwarfs in  
which H$_2$O lines were too weak (Paper II). However,
we estimate oxygen abundances in these four M dwarfs with the use of
log\,$A_{\rm O}/A_{\rm C}$ -- log\,$A_{\rm C}$ \footnote{We use the
notation $A_{\rm El} = N_{\rm El}/N_{\rm H}$, where $N_{\rm El}$ and
$N_{\rm H}$ are the number densities of the element El and hydrogen nuclei,
respectively.} relation based on  46 M dwarfs 
for which  carbon and oxygen abundances were determined as noted above.
A least square fit to log\,$A_{\rm O}/A_{\rm C}$ (data reproduced in 
eighth column of Table 1 and plotted in Fig.\,6b) is
$$ {\rm log}\,A_{\rm O}/A_{\rm C} = 0.270\,{\rm log}\,A_{\rm C}
           -0.684,   \eqno(1)  $$
from which we estimate log\,$A_{\rm O}/A_{\rm C}$ and then 
log\,$A_{\rm O}$ by the use of the known log\,$A_{\rm C}$.  
We include these four early M dwarfs with the estimated oxygen abundances
in our analysis of the carbon isotopic ratio based on CO spectra in 
this paper. We think that the use of the estimated oxygen abundance can be
acceptable, since CO spectra depend primarily on carbon abundance and 
effect of oxygen abundance is rather minor. 

From the  38
M dwarfs for which C and O abundances were determined, we exclude one
object GJ\,686  whose spectrum is disturbed by unknown noise. Also, we
could not include 2MASSI\,J1835379+325954 in our sample of eight late M 
dwarfs, since this object is a rapid rotator (Paper III) and isotopic 
features are completely smeared out. As a result, we have 37 + 7 + 4 = 48 
objects for our analysis of the carbon isotopic ratios in this paper. 

For 50 M dwarfs we have analyzed through Papers I to III,
object, spectral type, $T_{\rm eff}$, and log\,$g$ are given through
first to fourth column  in Table 1, and  
log\,$A_{\rm C}$, log\,$A_{\rm O}$ (including the oxygen abundances 
estimated with the use of equation 1), and log\,$A_{\rm O}/A_{\rm C}$ 
through sixth to eighth column.

\vspace{2mm}

------------------------------

table 1: p.16

------------------------------

\subsection{Model photospheres}
We apply dust-free model photospheres of C series included as a subset of 
our unified cloudy models (UCMs) for cool dwarfs \citep{Tsu02} throughout
this paper. Although we have prepared a grid of model photospheres\footnote{
http://www.mtk.ioa.s.u-tokyo.ac.jp/$\sim$ttsuji/export/ucm2},
we generated a specified model for each object based on $T_{\rm eff}$ and
log\,$g$ given in third and fourth columns of Table\,1, respectively
(Papers I \& III). The chemical composition assumed is based either  on 
the classical solar C and O abundances (\cite{And89}; referred to as 
{\it case a}; see Table\,1 of \cite{Tsu02}) or 
on the downward revised ones (\cite{All02}; {\it case c}), depending on the C 
and O abundances of each object in columns 6 and 7, respectively, in Table 1.  
The model we have applied is given in the fifth column of Table\,1 for
each object, and a model is designated as Ca or Cc/$T_{\rm eff}$/log\,$g$.
For example, Ca3570c489 implies a model of C series with the {\it case a}
chemical composition, $T_{\rm eff}$ = 3570\,K, and log\,$g$ = 4.89. 

\subsection{Molecular data}
We evaluate line positions of $^{13}$C$^{16}$O lines based on Dunham 
coefficients by \citet{Gue83}, and apply the same intensity data as for 
$^{12}$C$^{16}$O \citep{Cha83}.
As for spectroscopic data of H$_2$O, we continue to apply BT2-HITEMP2010 
database (\cite{Bar06}; \cite{Rot10}) as in Papers I - III.

The BT2-HITEMP2010 line list is largely based on the computed water line 
list, and its accuracy has been confirmed by extensive testings against 
astronomical 
and laboratory data by \citet{Bar06}. We also compare the line list with 
the observed water lines in sunspot spectra in the spectral region we are
to work. For this purpose, we use the Sunspot Umbral 
Spectrum \citep{Wal92} recorded with resolution as high as 
$R \approx 180000$ (FWHM $\approx 0.024$\,cm$^{-1}$). We find  14
undisturbed water lines between 4200 and 4240\,cm$^{-1}$ (or between 
23585 and 23810\,\AA) in the atlas and measured their line positions
as given in the second column of Table 2. For comparison, the line positions
from the BT2-HITEMP2010 database are given in third column. Since we are using
wavelength unit throughout this paper,  these values are converted to
the wavelength unit (in \AA\,  in vacuum) and given in fourth and fifth 
columns. The difference of the two is given in sixth column for each line.
(The upper and lower levels of the transition are given after seventh column.)
 Although the differences are biased towards negative, the differences 
themselves are rather small, less than 0.060\,\AA. On the other hand,
 the resolution of our M dwarf spectra is FWHM $\approx 16$\,km\,s$^{-1}$ 
or about 1.25\,\AA. Hence the accuracy of the computed water line list
is sufficient for our purpose. We again admire that the computed line
positions of water lines can be predicted with such an accuracy.

\vspace{2mm}

------------------------------

table 2: p.17

------------------------------

\newpage

\section{$^{13}$C features in M dwarfs }
The (2,0), (3,1), and (4,2) bands of the first overtone bands of 
$^{12}$C$^{16}$O show bandheads at 22935.2, 23226.6, and 
23524.6\,\AA\, (in vacuum), respectively,
and those of $^{13}$C$^{16}$O at 23448.3, 23739.4, and 24036.9\,\AA, 
respectively. The $^{13}$C$^{16}$O (2,0) and (3,1) bandhead regions
are well covered by our observed spectra. However, the (2,0) bandhead region
is blended with Ti I line at 23447.87\,\AA\, 
(a$^{3}$G$_{4}$ -- z$^{3}$F$_{3}^{0}$),
and this atomic line is pretty disturbing especially in early M dwarfs.
The (4,2) bandhead region of $^{13}$C$^{16}$O is not covered by our 
observed spectra. Then, we decide to use  the (3,1) bandhead region 
for our analysis of $^{13}$C$^{16}$O. 
     
Predicted spectra of the $^{13}$C$^{16}$O  (3,1) bandhead region
(23735 -- 23755\,\AA) for the $^{13}$C abundances of 1\% and 10\% of $^{12}$C
(or  $^{12}$C/$^{13}$C = 100 and 10, respectively)
are shown at the top panel of Fig.\,1 (Fig.\,1a). For this purpose, we apply
carbon and oxygen abundances together  with  model photosphere Ca3570c489 
for GJ\,15A, as an example (see Table 1). 
The spectra are first evaluated with a sampling interval of 0.02A
and then the results are convolved with the slit function (Gaussian) 
of FWHM = 16\,km\,s$^{-1}$. The micro-turbulent velocity is assumed to be
1\,km\,s$^{-1}$ throughout this paper. 
For comparison, a predicted spectrum of $^{12}$C$^{16}$O is shown in the
second panel of Fig.\,1 (Fig.\,1b). The bandhead region of $^{13}$C$^{16}$O
is well separated from the strong $^{12}$C$^{16}$O lines. A predicted
spectrum of H$_2$O in the same region is evaluated with the use of the
BT2-HITEMP2010 line-list (\cite{Bar06}; \cite{Rot10})
and shown in the third panel of Fig.\,1 (Fig.\,1c).  

Composite spectra of $^{13}$C$^{16}$O, $^{12}$C$^{16}$O, and H$_2$O
for $^{12}$C/$^{13}$C = 100 and 10 are shown in the bottom panel of
Fig.\,1 (Fig.\,1d). Features A and F (especially its red wing)
are little affected by the $^{13}$C$^{16}$O lines, and these features
are used as references in fitting observed and predicted spectra. 
 On the other hand,
features B, C, D, and E show the effect of $^{13}$C$^{16}$O lines,
and can be used to estimate  $^{12}$C/$^{13}$C  ratio.
Feature B shows the largest isotopic effect due to the contribution of
$^{13}$C$^{16}$O (3,1) bandhead, but this feature is also blended with
$^{12}$C$^{16}$O and H$_2$O lines (see Figs.\,1b and 1c). 
Feature C is dominated by H$_2$O lines \footnote{The major constituents
of feature C are H$_2$O lines nos. 6, 7, and 8 of Table 2, although
about 6000 weak H$_2$O lines are included in the H$_2$O spectrum
shown in Fig.\,1c.} except for early M dwarfs
(see Fig.\,1c), and cannot be a good indicator of isotopic effect.
On the other hand, feature D is dominated by the strong R\,13 line of 
$^{13}$C$^{16}$O (2,0) band at 23745.349\,\AA, and contributions of 
$^{12}$C$^{16}$O 
and H$_2$O lines are relatively minor (see Figs.\,1b and 1c). For
this reason, feature D can be an excellent indicator of $^{13}$C$^{16}$O.
Feature E is blended with  $^{12}$C$^{16}$O and H$_2$O lines, and
cannot be useful for our purpose. As for comparison with the observed 
spectrum of GJ\,15A, see Fig.\,2b.   

\vspace{2mm}

------------------------------

Fig. 1: p.10

------------------------------

\section{Results}

\subsection{Analysis on carbon isotopic ratios}

With our medium resolution spectra, it appears to be  rather difficult
to determine  $^{12}$C/$^{13}$C  by a visual comparison of the 
observed and predicted spectra.  Instead, we apply 
chi-square ($\chi^{2}$) analysis to judge goodness
of the fits between the observed and predicted spectra. 
For this purpose, we prepare predicted spectra  assuming
 $^{12}$C/$^{13}$C = 10, 25, 50, 100, and 200 
for the bandhead region of $^{13}$C$^{16}$O (3,1) band
with the physical parameters and abundances summarized in Table\,1.
The $\chi^{2}_{r}$ value for the fitting of the observed spectrum and 
predicted one for an assumed values of $^{12}$C/$^{13}$C 
is  evaluated from
$$ 
\chi^{2}_{r}(^{12}{\rm C}/^{13}{\rm C}) = {1\over{N-1}}\sum_{i=1}^{N}
{\Bigl(}{ {f_{\rm obs}^{i} - f_{\rm cal}^{i}(^{12}{\rm C}/^{13}{\rm C} ) }
\over\sigma_{i}}{\Bigr)}^{2},   \eqno(2) 
$$
where $f_{\rm obs}^{i}$ and  $f_{\rm cal}^{i}(^{12}{\rm C}/^{13}{\rm C} )$ 
are observed and predicted
(for a given $^{12}$C/$^{13}$C value) spectra normalized by their 
pseudo-continua, respectively, and $N$ is 
the number of data points. Also, $\sigma_{i}$ is the noise level  
assumed to be constant at 0.01 of the continuum level throughout \footnote{
The noise levels estimated from the S/N ratios given in the ninth column
of Table 3 are about 0.007 -- 0.02. We use a median value of 0.01
for all the objects for simplicity, since this value has no effect for
the minimization of $ \chi^{2}_{r}$  values in each object.}.
Then the resulting $ \chi^{2}_{r}$  values are
on relative scale, which is sufficient  for our purpose  to
find a best $^{12}$C/$^{13}$C value by the minimization
of the $\chi^{2}_{r}$  values for each object.

We apply the method outlined above to the observed 
spectra of  48 M dwarfs listed in Table 3.   Some examples are
shown  Figs.\,2a-l, where filled circles are observed spectra and 
solid lines are predicted ones for $^{12}$C/$^{13}$C
 = 10, 25, 50, 100, and 200 (from bottom to top in this order). 
In Figs.\,2a-l, the resulting 
$^{12}$C/$^{13}$C value (to be discussed below) is 
given at the upper left corner of each panel, following the
object's name and its spectral type.

We first notice in Figs.2a-l that feature D in observed spectra
of many objects show  a concavity, and this should be due to the 
$^{13}$C$^{16}$O (2,0) R\,13 line at 23745.349\,\AA, since  other 
contributions 
such as due to $^{12}$C$^{16}$O and/or  H$_2$O are not present or very weak
in this region as noted in section 3. Thus the concavity in feature D  
can be regarded as clear evidence for $^{13}$C$^{16}$O, and we are now
convinced the presence of a detectable amount of $^{13}$C$^{16}$O  in M 
dwarfs  for the first time.
Other $^{13}$C  indicator such as feature B also shows a concavity
but it is not sure if this is due to $^{13}$C$^{16}$O, since this
may also be due to $^{12}$C$^{16}$O and/or H$_2$O lines as noted
in section 3.

We then proceed to the $\chi^2$ analysis to find a best fit between the 
observed and predicted spectra. The resulting $\chi^{2}_{r}$ values for the 
comparisons of the observed spectrum and predicted spectra for five 
assumed values of  $^{12}$C/$^{13}$C = 10, 25, 50, 100, 
and 200  are given through  second to sixth column of Table 3 for each 
of our 48 program stars.
Then the minimization of $\chi^2_{r}$ is done by fitting a parabola
that passes through three points near the possible minimum on 
log\,$^{12}$C/$^{13}$C  -- $\chi^2_{r}$  plane for each object. Some
examples are shown in Fig.\,3 for the cases of GJ\,876 ($^{12}$C/$^{13}$C
 = 39), 849 ($^{12}$C/$^{13}$C = 67), 212 ($^{12}$C/$^{13}$C = 134), and 
526 ($^{12}$C/$^{13}$C $> $200), and the resulting $^{12}$C/$^{13}$C value
corresponding to the minimum of $\chi^2_{r}$ for each objects
is given in  seventh column of Table 3. Also, the corresponding value of
log\,$^{13}$C/$^{12}$C = $-$log\,$^{12}$C/$^{13}$C
is given in eighth column of Table 3 for each object.

From Table 3, we know that some objects show rather low $^{12}$C/$^{13}$C
values (i.e., $^{13}$C rich) and  values of $^{12}$C/$^{13}$C are
found to be 39, 44, and 49 in GJ\,876, 15A, and 179, respectively.
These are the most $^{13}$C rich cases in our sample, and we could not 
find more $^{13}$C rich cases such as found in some evolved high 
luminous cool stars. Inspection of Figs.\,2a-c reveals that feature D
clearly shows the concavity and we are convinced $^{13}$C rich nature of
these objects. Overall fits such as in features B and C
between observed and predicted spectra also appear reasonable for the 
case of  $^{12}$C/$^{13}$C$\approx 50$ in Figs.\,2a-c (remember 
that the predicted spectra shown by the solid lines
are assuming  $^{12}$C/$^{13}$C = 10, 25, 50, 100, and 200 from
bottom to top in each panel).

The $\chi^2$ analysis for GJ\,324B  suggests  $^{12}$C/$^{13}$C = 56, 
and features B and D are roughly consistent with this result  on Fig.\,2d.
However, feature C appears to suggest $^{12}$C/$^{13}$C$\approx 25$.
But feature C is badly disturbed by H$_2$O lines and such inconsistency
may be fatal for the medium resolution spectra we are using.
The $\chi^2$ analysis on GJ\,849  suggests  $^{12}$C/$^{13}$C = 67, 
and overall fits are roughly consistent with this result  on Fig.\,2e.
Although the $\chi^2$ analysis for GJ\,231.1B  suggests 
$^{12}$C/$^{13}$C 
= 70 (about the same as for GJ\,849), it is more difficult to confirm 
the overall fits on Fig.\,2f. Even though the spectral types of
 GJ\,849 and GJ\,231.1B are  the same, GJ\,849 is hotter by about
100\,K and more carbon rich compared with GJ\,231.1B (see Table\,1).
For this reason, H$_2$O lines are weaker and CO lines are stronger in GJ\,849
than in GJ\,231.1B, making it easier to investigate isotopic effect 
on CO spectra in GJ\,849.   

We find  $^{12}$C/$^{13}$C values to be 80, 90, and 117 in
GJ\,229, HIP\,79431 and GJ\,649, respectively,  from the  $\chi^2$ analysis.
Inspection of Figs.\,2g-f reveals that feature D in all these three
cases clearly shows the concavity and the presence of $^{13}$C$^{16}$O in 
these objects are well demonstrated even for the rather large 
$^{12}$C/$^{13}$C values (i.e., low $^{13}$C abundances). However, to see 
the overall fits in features B, C, and D are more difficult since the effect 
of $^{12}$C/$^{13}$C on the predicted spectra tends to be minor for large 
$^{12}$C/$^{13}$C values. 
We must recognize that it is increasingly difficult to find the best fit
by the visual comparison of the observed and predicted spectra and,
for this reason, we adopt  the $\chi^2$ analysis which is  more robust
for our purpose. 

Such a difficulty in the cases of large $^{12}$C/$^{13}$C values is more 
severe in GJ\,212 for which the $\chi^2$ analysis suggests 
$^{12}$C/$^{13}$C = 134, but it is difficult to confirm this result by 
the visual inspection of Fig.\,2j in which predicted spectra for 
$^{12}$C/$^{13}$C = 100 and 200 differ very little. For GJ\,205 shown in 
Fig.\,2k, the $\chi^2$ analysis suggests $^{12}$C/$^{13}$C = 140.
Even for such a high value of $^{12}$C/$^{13}$C, feature D clearly shows 
the concavity and
hence  presence of $^{13}$C$^{16}$O. Although this object is classified 
as dM3, this does not necessarily imply low temperature but this
should largely be due to the high metallicity of this object as noted in
Paper I. In fact, this object is the most carbon and oxygen rich in our 
sample (see Table\,1). For this reason, CO lines appear to be
enhanced and $^{13}$C$^{16}$O lines as well. Also, H$_2$O lines, which are
disturbing for investigating $^{13}$C$^{16}$O lines,  are not 
so strong due to higher temperature than that suggested for the spectral
type dM3. As a result, this object provides a favorable case for the
study of the carbon isotopic ratio. Finally, the $\chi^2$ analysis suggests 
$^{12}$C/$^{13}$C\,$> 200$  for GJ\,526 and this result is well consistent 
with Fig.\,2l. For example, feature D does not show the concavity at all, 
indicating small abundance of $^{13}$C$^{16}$O. More or less similar 
results of $^{12}$C/$^{13}$C\,$> 200$ are obtained for other 16 M dwarfs 
(e.g., Figs.\,4b and 4c).

In principle, it is desirable that the results of the $\chi^2$ analysis
could be confirmed to be consistent with the visual comparison of
observed and predicted spectra. As we have noticed above, 
this is difficult especially for the cases of large $^{12}$C/$^{13}$C
values, in which  predicted spectra of $^{13}$C$^{16}$O are already very 
weak and a difference for the different $^{12}$C/$^{13}$C values 
should  be still smaller. However, in our $\chi^2$ analysis, we have 
changed only $^{12}$C/$^{13}$C values
keeping all the other parameters unchanged. Hence the changes of  the
$\chi^2_r$ values for changed $^{12}$C/$^{13}$C  values should reflect the
effect of $^{12}$C/$^{13}$C alone, and  the  
$^{12}$C/$^{13}$C value minimizing $\chi^2_{r}$ should provide the best fit.
We must, however, keep in mind that the resolution of our spectra is
by no means sufficient for our analysis outlined in this section.
For this reason, our result is only very preliminary and we hope that 
the problem of the carbon isotopic ratios in M dwarfs will be pursued 
further by higher resolution spectroscopy in the future. 

\vspace{2mm}

------------------------------

Fig. 2: p.11

Fig. 3: p.12

\vspace{2mm}

table 3: p.18

------------------------------

\subsection{A binary test}

Our sample includes three binary pairs, GJ\,338A and 338B,
GJ\,725A and 725B, and GJ\,797B-NE and 797B-SW. Since abundances
including isotopic ratios may not be different for the objects
in a binary pair \footnote{In fact, the carbon and oxygen abundances in these
binary pairs agree within 0.03\,dex (note that this is smaller than the 
probable error of each abundance determination) in all the three cases, as can 
be confirmed in Table 1.}, we hope that an assessment of accuracy of
our result can be obtained by comparing the results in these binary pairs. 

The results of the $\chi^2$ analysis reveal that $^{12}$C/$^{13}$C = 105 and   
$^{12}$C/$^{13}$C\,$> 200$ for GJ\,338A and 338B, respectively 
(see Table 3). Comparisons
of the observed and predicted spectra are shown in Figs.4a and 4b for 
GJ\,338A and 338B, respectively. The results shown in Figs.\,4a-b are 
not much different. But observed and predicted spectra in feature C
show some differences in GJ\,338A and in GJ\,338B, and
such a difference may be an origin of the different results on
$^{12}$C/$^{13}$C  by the $\chi^2$ analysis. However, the predicted spectra 
for $^{12}$C/$^{13}$C = 100 and 200 are almost indistinguishable in feature C
and it should be difficult to decide which of these $^{12}$C/$^{13}$C values 
should apply to these M dwarfs. Thus, we must accept uncertainty  of about 
a factor of two in our $\chi^2$ analysis if $^{12}$C/$^{13}$C\,$> 100$.   

The results of the $\chi^2$ analysis reveal that $^{12}$C/$^{13}$C\,$> 200$ 
and $^{12}$C/$^{13}$C = 135 for GJ\,725A and 725B, respectively. Comparisons
of the observed and predicted spectra are shown in Figs.4c and 4d
for GJ\,725A and 725B, respectively. Again, the results for the 
two objects are not
much different, and to decide which of $^{12}$C/$^{13}$C\,$\approx 100$ or 
200 should apply to each of these objects appears to be difficult,
since the predicted spectra for $^{12}$C/$^{13}$C = 100 and 200 differ little.
However, our result of the $\chi^2$ analysis, which is a kind
of statistical analysis, suggests $^{12}$C/$^{13}$C\,$> 100$  both 
for GJ\,725A and 725B, and we again find  consistency within about a factor
of two in our analysis.

The results of the $\chi^2$ analysis reveal that $^{12}$C/$^{13}$C = 116 and   
$^{12}$C/$^{13}$C = 93 for GJ\,797B-NE and 797B-SW, respectively. Direct 
comparisons
of the observed and predicted spectra are shown in Figs.\,4e and 4f for
GJ\,797B-NE and 797B-SW, respectively. In this case, the $\chi^2$ 
method gives the results that agree rather well for this binary pair.
Although different features (e.g., B, C, and D) in Fig.\,4e and 4f do 
not necessarily indicate the same $^{12}$C/$^{13}$C values by visual 
comparisons,
the $\chi^2$ values in Table 3 show the integrated effect of $\chi^2$ 
values for different features such as B, C, D and E. For this reason, our 
$\chi^2_{r}$ values in Table 3 can be a measure of overall fits, and 
our $\chi^2$ analyses for GJ\,797B-NE and 797B-SW show  reasonable
consistency in the overall fits if not in individual features.

The above results imply that it is difficult to determine $^{12}$C/$^{13}$C
value when it is as large as 200, and if the result of the $\chi^2$ analysis 
reveals such a result as $^{12}$C/$^{13}$C\,$> 200$, it is still possible 
that actual $^{12}$C/$^{13}$C  may be as low as 100 (e.g., GJ\,338B and 
GJ\,725A). Certainly,
it is natural that our $\chi^2$ analysis does not work for such small 
difference which our observed data could not resolve. However, if 
$^{12}$C/$^{13}$C\,$\approx 100 $, our binary test suggests that the results
agree rather well as in the  case of GJ\,797B.  Although our 
sample of binary pairs is not large enough,
we regard our result of $^{12}$C/$^{13}$C\,$> 100$ to be uncertain by 
about a factor of two but should be better in the case of 
$^{12}$C/$^{13}$C\,$\lesssim 100$.

\vspace{2mm}

------------------------------

Fig. 4: p.13

------------------------------

\subsection{Further test on the accuracy}

Unfortunately, our sample of binary is limited to three pairs and to the cases
of rather high  $^{12}$C/$^{13}$C ratios. To extend the similar test
to other cases, we use the spectra recorded at A and B positions
on slit alternatively during the observation. The resulting
spectra are co-added at the end and used for our analysis so far done
(Papers I, II, and III; subsections 4.1 and 4.2 in Paper IV). These two 
spectra observed at slit positions A
and B are recovered separately and we analyze these spectra A and B
as if they are independent two spectra. Although S/N ratios of these
spectra are reduced by about a factor of $\sqrt{2}$, 
these spectra can be used
in the same way as in the binary test outlined in the previous subsection.

For this purpose, we select M dwarfs of various $^{12}$C/$^{13}$C ratios
and  we choose GJ\,876, 849, 212, and 526 as examples.
These objects cover the $^{12}$C/$^{13}$C ratios from the smallest to
the largest and, for this reason, these objects are already used as 
representative cases in Fig.3. The spectra observed at A and B positions are
analyzed by essentially the same way as we analyze
the co-added spectrum in subsection 4.1  and the results 
are shown in Figs.\,5a-d. The $\chi^{2}_{r}$ values for  the fitting of
the  spectrum observed at  A or B position  
and the predicted spectra  for five assumed values of
the $^{12}$C/$^{13}$C ratio evaluated by equation\,(2) are given in 
second through sixth columns of Table\,4. The result of the minimization
of $\chi^{2}_{r}$ values is given at the last column of Table 4 for
each spectrum observed at positions  A and B.   

The results for the case of GJ\,876, which is 
the most $^{13}$C rich in our sample, reveal that the resulting 
$^{12}$C/$^{13}$C ratios are 37.7 and 31.4 for the spectra observed at
A and B positions (hereafter be referred to as spectra A and B), respectively.
Thus, the results for the spectra A and B agree rather well and this may 
be due to small differences in the features B and D which are most sensitive 
to the $^{12}$C/$^{13}$C ratio (Fig.\,5a). On the other hand, the resulting 
$^{12}$C/$^{13}$C ratios are 59.3 and 108.3 for the spectra  A and B, 
respectively, 
in the case of GJ\,849, and the difference of the $^{12}$C/$^{13}$C ratios 
based on the spectra  A and B is rather large. This may be due
to a considerable difference of the feature D in the observed spectra A and B
(Fig.\,5b). The S/N ratios for GJ\,876 and 849 are not much different 
(see Table 3), but it appears that the effect of noise on the spectra
is rather different and complicated. The resulting $^{12}$C/$^{13}$C ratios 
are 106.9 and 139.7 for the spectra A and B, respectively, 
in the case of GJ\,212. The difference of the $^{12}$C/$^{13}$C ratio based 
on the spectra  A and B is rather modest. In this case, the $\chi^{2}_{r}$ 
values indicates the $^{12}$C/$^{13}$C ratios not very different, despite 
some differences in the spectra observed at positions A and B (Fig.\,5c). 
In fact, even if the difference of the observed and predicted spectra 
(i.e., $f_{\rm obs}^{i} -　f_{\rm cal}^{i}$) at a particular wavelength
is disturbed by noise,  the $\chi^{2}_{r}$ value is determined 
by including the contributions from other wavelength regions less
disturbed by noise as well, and the effect of noise can be averaged
in the resulting $\chi^{2}_{r}$ value.
Finally, the resulting $^{12}$C/$^{13}$C ratios for the spectra
observed at A and B positions for GJ\,526 are larger than 200 and 141.0,
respectively. Here, the situation is quite similar to     the binary
pairs GJ\,338A,B and GJ\,725A,B discussed in subsection 4.2: Since
the effect of $^{13}$C$^{16}$O lines on the spectrum is difficult to
discriminate if $^{12}$C/$^{13}$C $ \gtrsim 200 $ it is difficult to
determine $^{12}$C/$^{13}$C ratio with the medium resolution we
are using.

The results outlined above are summarized in Table\,5: For each
object, the $^{12}$C/$^{13}$C ratios based on the spectra observed at
A and B positions are shown in second and third columns, respectively.
The mean value of the $^{12}$C/$^{13}$C ratios based on the
spectra at A and B positions is given in fourth column, and 
the probable error is also  estimated. The resulting probable errors
appear to be not so large: about 10\% in  favorable cases (GJ\,876,
212) and about 30\% in a less favorable case (GJ\,849). But uncertainty 
of as large as a factor of two must be accepted in the poor $^{13}$C cases
(e.g., GJ\,526).   The $^{12}$C/$^{13}$C ratio based on the co-added 
spectrum is reproduced in fifth column from Table 3. Although an
error cannot be estimated from a single spectrum alone, the resulting
$^{12}$C/$^{13}$C ratio is roughly within the error bar estimated 
in the fourth column by using the spectra with S/N ratios reduced 
by a factor of $\sqrt{2}$. Thus, we may conclude that the accuracy
of our results on the $^{12}$C/$^{13}$C ratios can be acceptable, and
at least better than 50\% \footnote{By the way, the $^{12}$C/$^{13}$C ratio
of the binary system GJ\,797B can be
 104.5 $\pm$ 11.0 as a mean of those of its constituents GJ\,797B-NE 
and 797B-SW discussed in subsection 4.2, and again the error of the 
resulting $^{12}$C/$^{13}$C ratio can be estimated to be about 10\%.} 
except for the cases of  the poor $^{13}$C (e.g., $^{12}$C/$^{13}$C 
$ \gtrsim 200$). Although our spectral resolution is not high enough to 
resolve the individual $^{13}$C$^{16}$O lines and the S/N ratios are not
very high, the $\chi^{2}_{r}$ value reflects the integrated effect in the
difference of observed and predicted spectra. For this reason, the 
$\chi^{2}$ analysis can be a powerful tool for assessment of 
the goodness of the fitting.

\vspace{2mm}

------------------------------

Fig. 5: p.14

\vspace{2mm}

table 4: p.19

table 5: p.19

------------------------------

\section{Discussion}

The unweighted mean value of 48 M dwarfs including those with the lower 
limit of 200 is $ (^{12}{\rm C}/^{13}{\rm C})_{\rm dM} > 127 \pm 41$ (p.e.). 
If we exclude such a case for
which only the lower limit of 200 is known, the unweighted mean value of 
31 M dwarfs is $ (^{12}{\rm C}/^{13}{\rm C})_{\rm dM} = 87 \pm 21$ (p.e.).
This value agrees well with the value of the solar system, 
$ (^{12}{\rm C}/^{13}{\rm C})_{\rm SS} = 89.9$ \citep{And89}. Also, the 
 value of the solar
photosphere based on the high resolution infrared solar spectrum is 
$ (^{12}{\rm C}/^{13}{\rm C})_{\rm Sun} = 84 \pm 5$ \citep{Har87} or 
$ (^{12}{\rm C}/^{13}{\rm C})_{\rm Sun} = 80 \pm 1$ \citep{Ayr06}. 
Then the $^{12}$C/$^{13}$C ratio of the solar photosphere as well as of
the solar system is  typical of the unevolved late-type dwarfs.
However, the scatter of the $^{12}$C/$^{13}$C ratios in Table 3 
is rather large extending from $^{12}{\rm C}/^{13}{\rm C}$ = 39 to 
$ ^{12}{\rm C}/^{13}{\rm C} > 200$ (also see Fig.\,6a). 

The resulting  $^{12}$C/$^{13}$C values in Table 3 are rewritten as
log\,$^{13}$C/$^{12}$C in eighth column, and plotted against 
log\,$A_{\rm C}$ in Fig.\,6a, where log\,$^{13}$C/$^{12}$C 
shows a large scatter without clear 
dependence on the  metallicity. This result shows a marked contrast to the
plot of log\,$^{16}$O/$^{12}$C (= log\,$A_{\rm O}/A_{\rm C}$ from eighth
column of Table 1) 
against log\,$A_{\rm C}$, which shows a rather tight correlation
with a higher  ratio at the lower metallicity, as shown in  Fig.\,6b
(reproduced from Fig.\,19  in Paper III). Such a contrast between 
Figs.\,6a and 6b  can be a natural consequence that $^{16}$O and 
$^{12}$C are the primary products in the stellar nuclear synthesis 
while $^{13}$C is the secondary product, at least partly (e.g., 
\cite{Ren81}). Thus the difference of the primary and secondary 
products in the actual distribution of the isotopes is clearly 
demonstrated for the first time.

The $^{12}$C/$^{13}$C ratios of late-type dwarfs should reflect
those of the ISM from which the stars are formed. For comparison,
the $^{12}$C/$^{13}$C ratio in the ISM have extensively been studied by
the millimeter spectroscopy of giant molecular clouds. 
The initial results reviewed by \citet{Wan80} show that there is no evidence
for a $^{12}$C/$^{13}$C gradient in the Galactic disk and the
 mean $^{12}$C/$^{13}$C ratio is $ (^{12}{\rm C}/^{13}{\rm C})_{\rm ISM} = 
60 \pm 8$, except for
the Galactic center source showing $^{12}$C/$^{13}$C $\approx 20$.
However, later results reviewed by \citet{Wil94} clearly indicated
an increase of the $^{12}$C/$^{13}$C ratio with the Galactocentric distance
 $D_{\rm GC}$:  The $^{12}$C/$^{13}$C ratio is 
$ (^{12}{\rm C}/^{13}{\rm C})_{\rm ISM} \approx 50 \pm 8$ at $D_{\rm GC} 
\approx$ 4\,Kpc and $ (^{12}{\rm C}/^{13}{\rm C})_{\rm ISM} \approx 
76 \pm 7$ at $D_{\rm GC} \approx$ 8.5\,Kpc (near the Sun). More recently,
the $^{12}$C/$^{13}$C ratios in the local ISM (within about 1 Kpc of the Sun)
have been determined from the optical spectroscopy of $^{13}$CH$^{+}$
absorption lines towards early type stars (\cite{Cas05}; \cite{Sta08}) and 
a significant scatter of 
$ (^{12}{\rm C}/^{13}{\rm C})_{\rm ISM} \approx$ 44 -- 147 was found in 
the local ISM. 

Our result on the $^{12}$C/$^{13}$C ratios in M dwarfs (Fig.\,6a) shows more 
or less similar pattern to those in the ISM outlined above. For example,
the scatter of the $^{12}$C/$^{13}$C ratio is quite large and extends from 
$\approx 40$ to $ > 100$ both in the ISM and in M dwarfs. Also, such a low 
ratio as below 40 is not found both in the ISM and in
M dwarfs \footnote{It was suggested that $^{12}$C/$^{13}$C $>15$ for
GJ\,752B (VB10) by \citet{Pav02}, using spectra of 
$\lambda/\Delta\,\lambda = 1085$ or velocity resolution of 
276\,km\,s$^{-1}$. However, their resolution may be too
low to justify their result, as the authors themselves noticed.}.  
However, the  results on the ISM are on the present ISM at the 
present locations while our results on M dwarfs may  reflect the
$^{12}$C/$^{13}$C ratios  of the ISM at various ages  when M dwarfs 
were formed at different  birthplaces. 
For this reason, interpretation of the $^{12}$C/$^{13}$C ratios in M 
dwarfs should be more complicated than that in the ISM.

As to the origin of $^{13}$C, it is generally thought that $^{13}$C is 
formed from  $^{12}$C originally present in the star at the time of its 
formation, and hence $^{13}$C is regarded as a typical secondary element. 
Then, $^{13}$C can be produced in almost in all the stars in their 
hydrogen burning phase and will be dredged-up
to the surface by the convective zone developed in the red giant phase
(and eventually to ISM by stellar mass loss). However, the temperature
at the base of the convective envelope after He shell flash can be high
enough for CNO cycle reactions take place (hot bottom burning - HBB), 
and  $^{12}$C produced by the  He burning can also be converted into $^{13}$C. 
This $^{13}$C is primary since it comes from the primary $^{12}$C 
produced from He anew in the star \citep{Ren81}. For this reason, the origin 
of $^{13}$C in the Galaxy is by no means well settled. The chemical 
evolution of $^{12}$C/$^{13}$C in the Galaxy has been studied by 
assuming a mixed 
origin of primary and secondary (case A) or a purely secondary origin 
(case B) \citep{Pra96}. Their results show that the $^{12}$C/$^{13}$C 
ratio decreases  both in cases A and B during the evolution of
the Galaxy, and case A appears to explain somewhat better the present-day
$^{12}$C/$^{13}$C ratio if it is $(^{12}{\rm C}/^{13}{\rm C})_{\rm ISM}
\approx {70 \pm 10}$.  

Our result that the mean value of the $^{12}$C/$^{13}$C ratios 
in M dwarfs is somewhat larger than that of the present ISM, 
especially if we consider our result that many M dwarfs show 
$^{12}$C/$^{13}$C $> 200$, is well consistent with  the
 decrease of $^{12}$C/$^{13}$C ratios as a result of chemical
evolution of the ISM outlined above. 
Our result shown in Fig.6 suggests that $^{13}$C may be produced
as the secondary product, at least partly, and
since the secondary process depends 
on many factors including  nuclear reactions, mixing, mass-loss, binarity 
etc in evolved stars, distribution of $^{12}$C/$^{13}$C ratios in
the ISM and hence in M dwarfs should be inhomogeneous. However, the mean
$^{12}$C/$^{13}$C ratio should decrease with time in the ISM and hence 
in M dwarfs, resulting in a larger mean $^{12}$C/$^{13}$C ratio in M dwarfs
than that in the present ISM.

The  $^{12}$C/$^{13}$C ratio in M dwarfs in Table 3 and shown in Fig.\,6a
should reflect the $^{12}$C/$^{13}$C ratios of the ISM from which M dwarfs 
were formed, and the $^{12}$C/$^{13}$C ratio of the ISM itself has been 
determined by the mass-loss from the stars of the previous generation. It is
not possible to know the $^{12}$C/$^{13}$C ratios of the stars
of the previous generation, but we can study the $^{12}$C/$^{13}$C 
ratios of similar stars that might have contributed to the 
$^{12}$C/$^{13}$C ratios of M dwarf stars. For this purpose,
the $^{12}$C/$^{13}$C ratios in red giants, supergiants, 
and AGB stars that have been studied by many authors
as noted in section\,1 can be of some help. For example, if
relatively large production of carbon in metal-rich era (Fig.\,6b)
is due to ABG stars (e.g., \cite{Nom13}), the $^{12}$C/$^{13}$C ratios
in unevolved stars as well as in the ISM
may be related to those of AGB stars. However, it is not likely that  
the $^{12}$C/$^{13}$C ratios observed in the present evolved stars
can explain well  the  $^{12}$C/$^{13}$C ratio observed in the present
ISM as well as in M dwarfs. This should be because the  $^{12}$C/$^{13}$C 
ratio could be determined only from molecular spectra and the isotopic ratios
of many other objects not showing molecular spectra remain unexplored. 
For this reason, our knowledge on the distribution of isotopes in the 
Universe is quite limited  and imperfect. Unfortunately, this fact makes 
it difficult to understand observationally  the
origin of the carbon isotopes in M dwarfs as well as in the ISM. 
 
Other sources of information on the isotopic composition
in the Universe are primitive meteorites containing grains of silicon
carbide, graphite, diamond etc. Direct analysis on these grains provides
accurate isotopic compositions of celestial objects that have produced the
grains, and it is no longer limited by the observability of molecular
spectra. For example, $^{12}$C/$^{13}$C ratio obtained from interstellar
graphite shows a wide range from near the equilibrium value of the CN cycle
to several thousands (e.g., \cite{Ott93}; \cite{Ama93}). Clearly, the
dynamic range of the $^{12}$C/$^{13}$C ratio in meteorite sample is
much larger than in ISM and in M dwarfs. We hope that these
results can be incorporated into our understanding on the chemical
evolution of isotopes, although it may be by no means easy to identify the 
origin of each grain. 

\vspace{2mm}

------------------------------

Fig. 6: p.15

------------------------------

\section{Concluding remarks}

We have tried to estimate the carbon isotopic ratio  $^{12}$C/$^{13}$C
in M dwarfs based on the spectra of medium resolution 
($\lambda/\Delta\,\lambda \approx 20000$). Although this resolution
is certainly not sufficient to resolve faint $^{13}$C$^{16}$O lines
from the stronger lines of $^{12}$C$^{16}$O and H$_2$O, we find
clear evidence for the $^{13}$C$^{16}$O feature in the spectra of M 
dwarfs for the first time (see Fig.\,1 and Figs.\,2a - l). Then, 
determination of the $^{12}$C/$^{13}$C
ratios in M dwarfs is quite feasible especially if higher spectral
resolution can be employed, and we hope that the isotopic analysis
on low luminosity stars such as M dwarfs will be done more intensively
as in high luminosity cool stars. 

With our medium resolution, we determine a possible best value of
$^{12}$C/$^{13}$C in each M dwarf by comparing observed and predicted spectra,
not by a direct inspection but by an application of the  chi-square method.
Although the direct comparisons (see Figs.\,2a - l, 3a-f) show difficulties 
inherent to the spectra of insufficient resolution, the chi-square
analysis works reasonably well (see Fig.\,3). As a result, we determine
$^{12}$C/$^{13}$C ratios for 31 M dwarfs and a lower limit of 200 for
17 M dwarfs. From our preliminary survey, we may suggest that the
$^{12}$C/$^{13}$C ratios in M dwarfs are larger than about 40 and  not
likely to be so small as in some evolved high luminosity stars. 
Although the lower limit of the $^{12}$C/$^{13}$C ratio is found to be
200 in many M dwarfs, this result is uncertain by a factor of two or so.
Unfortunately, it is beyond the capability of our medium resolution spectra
to analyze faint features due to very low abundance of $^{13}$C.
We hope that our result can be improved by the
use of  higher spectral resolutions in the near future.
  
The mean $^{12}$C/$^{13}$C ratio in M dwarfs turns out to be
larger than that of the ISM (section\,5).
While the $^{12}$C/$^{13}$C ratio in the ISM provides the present
$^{12}$C/$^{13}$C ratio, that in M dwarfs reflects the carbon isotopic 
ratio  in the past ISM.  So far, the solar system, conserving
the $^{12}$C/$^{13}$C ratio of the ISM 4.5 Gyr ago, was used as a reference
in investigating
the time variation of the $^{12}$C/$^{13}$C ratio in the ISM, but M dwarfs 
can in principle be used for the same purpose with numerous objects in a 
larger time-space domain. In fact, M dwarfs offer an unique possibility
to determine  the $^{12}$C/$^{13}$C ratios in unevolved late-type dwarfs
and should have important role to trace the evolution
of carbon isotopes (and possibly other isotopes)  in the Galaxy. For
this reason, determination of isotopic abundances in M dwarfs should
be the next major challenge when high resolution infrared spectroscopy
will be ripe enough to be able to observe many faint objects not well
observed so far.

\bigskip
   I  thank T. Nakajima for recovering the spectra observed at slit
positions A and B before co-adding and for helpful discussion on the 
evaluation of the S/N ratio. I also thank him  and Y. Takeda for sharing 
the spectra of M dwarfs observed at Subaru IRCS.

 I thank the anonymous referee for careful reading of the text and for
critical comments especially on the accuracy of the analysis. 

  This research has made use of the SIMBAD operated at CDS, Strasbourg, France.
  
  Computations are carried out on common use data analysis computing system at 
the Astronomical Data Center, ADC, of the National Astronomical Observatory 
of Japan.

\newpage
\onecolumn

\begin{figure}
   \begin{center}
       \FigureFile(80mm,0mm){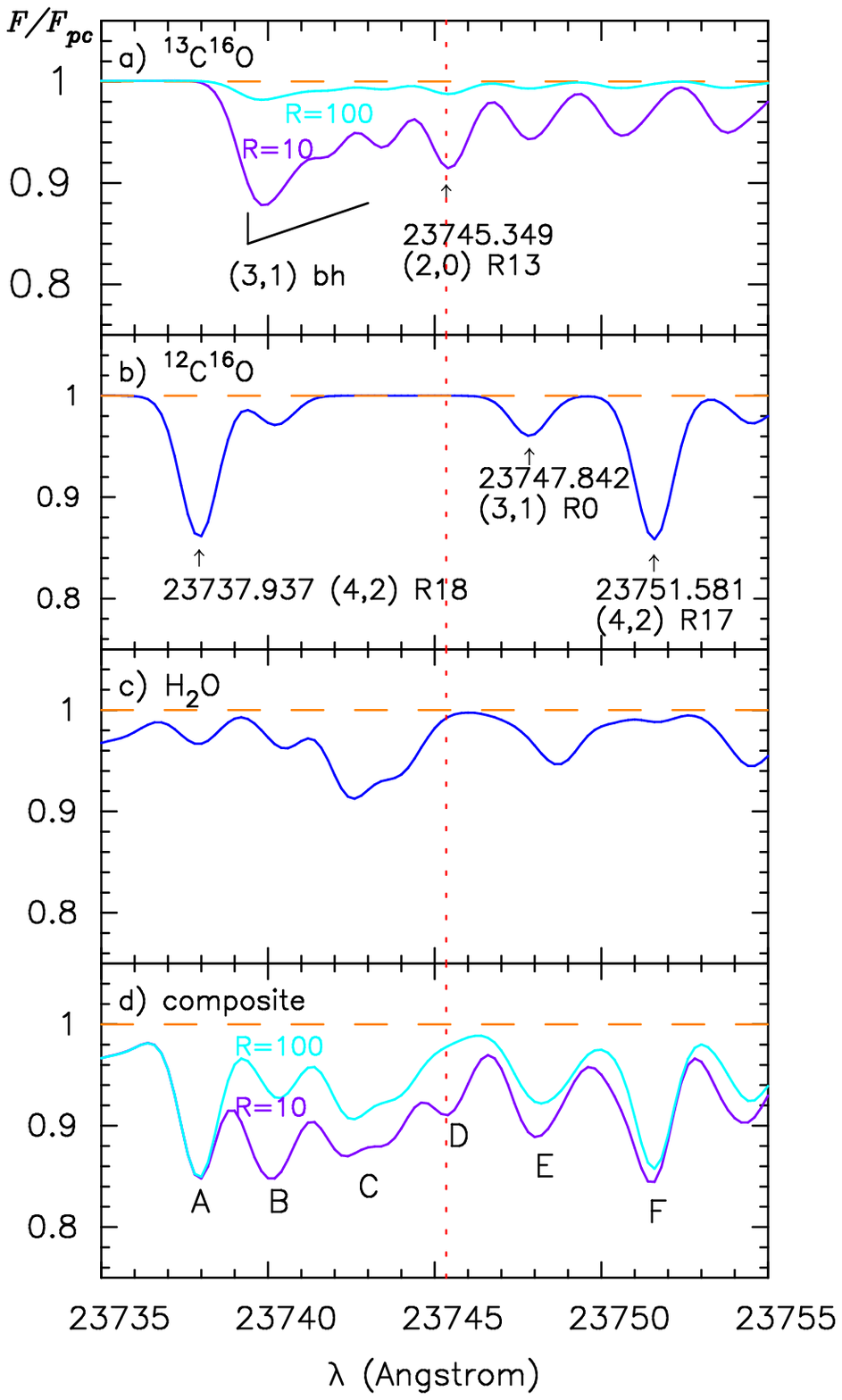}
   \end{center}
   \caption{
Model spectra near the bandhead region of $^{13}$C$^{16}$O (3,1) band based on 
log\,$A_{\rm C} = -3.60$, log\,$A_{\rm O} = -3.31$, and Ca3570c489: 
a) for $^{13}$C$^{16}$O alone, assuming $^{12}$C/$^{13}$C = R = 10 and 100. 
b) for $^{12}$C$^{16}$O alone. 
c) for H$_2$O alone.
d) for composite of $^{13}$C$^{16}$O (assuming $^{12}$C/$^{13}$C = R = 10 
and 100), $^{12}$C$^{16}$O, and H$_2$O. Features A and E (red wing) are
mostly due to $^{12}$C$^{16}$O. Feature B is dominated by $^{13}$C$^{16}$O
(3,1) bandhead but disturbed by  $^{12}$C$^{16}$O and H$_2$O lines. Feature C 
is dominated by H$_2$O lines except for early
M dwarfs. Feature D is dominated by $^{13}$C$^{16}$O (2,0) R13 line
without serious blending with $^{12}$C$^{16}$O and H$_2$O lines, and
can be an excellent indicator of $^{13}$C$^{16}$O.   }
\label{figure1}
\end{figure}

\begin{figure}
   \begin{center}
       \FigureFile(155mm,0mm){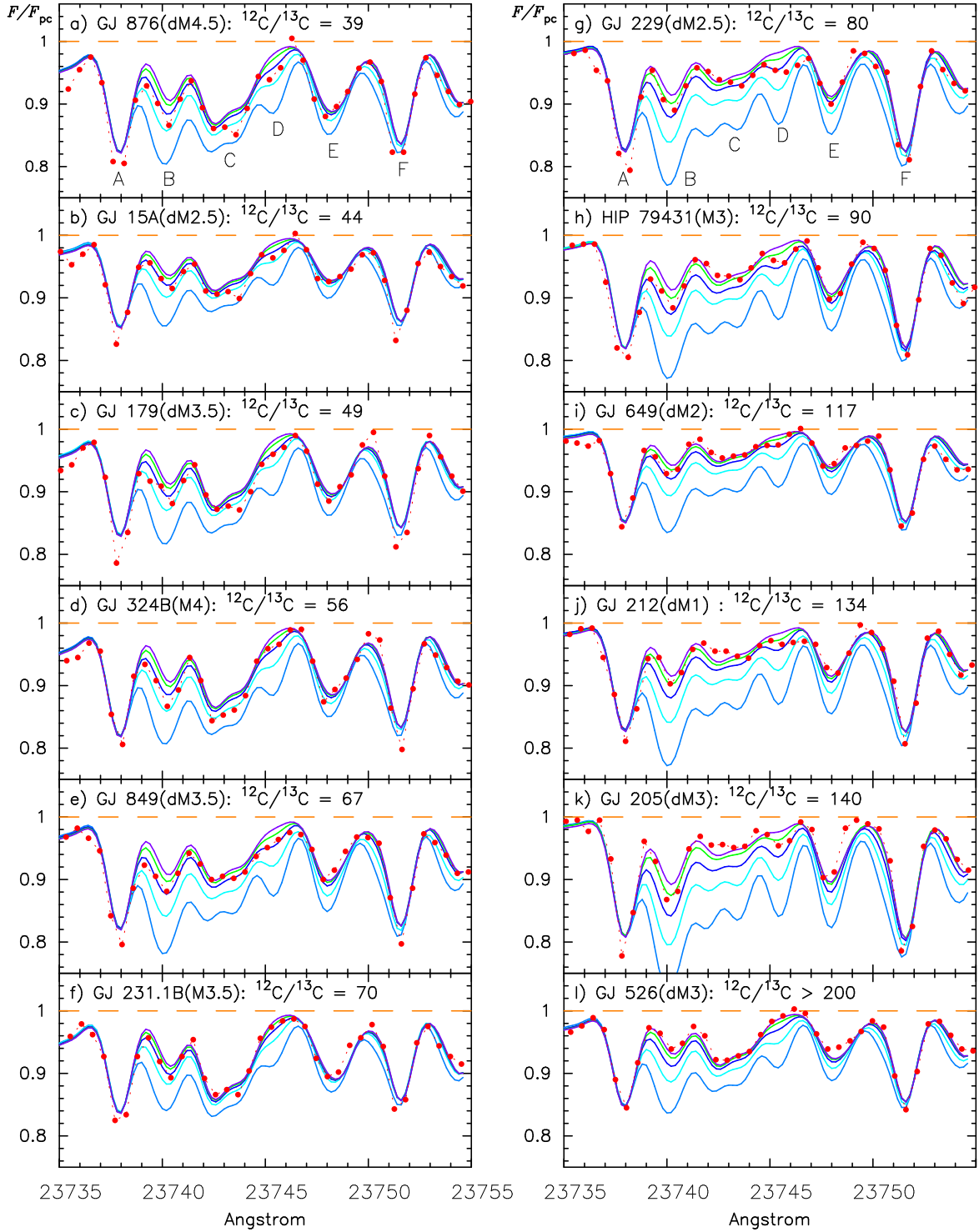}
   \end{center}
   \caption{
    Comparisons of the observed spectrum (filled circles) and 
predicted spectra for $^{12}$C/$^{13}$C = 10, 25, 50, 100, and 200
(solid lines from bottom to top in this order) in each panel. The object 
identification, spectral type, and resulting $^{12}$C/$^{13}$C value 
are indicated at the upper left corner for each object.
}
\label{figure2}
\end{figure}

\begin{figure}
   \begin{center}
       \FigureFile(75mm,0mm){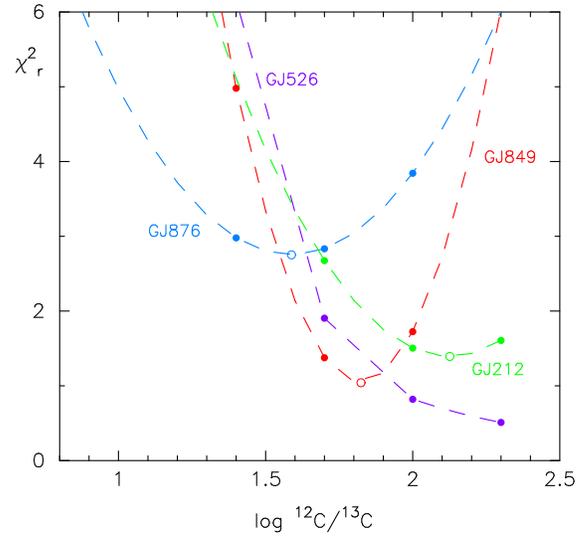}
   \end{center}
   \caption{
    The $\chi^{2}_r$ values given in Table\,3 are plotted against 
 log\,$^{12}$C/$^{13}$C for GJ\,876, 849, 212, and 526, as examples. 
A parabola is fitted to  three data points (filled circles) near the 
possible minimum 
and minimized $\chi^{2}_r$ value is shown by an open circle for each object.
However, the $\chi^{2}_r$ value for GJ\,526 shows no minimum and 
continues to decrease until $^{12}$C/$^{13}$C = 200. Then, we can 
determine only the lower limit of $^{12}$C/$^{13}$C to be 200.
}
\label{figure3}
\end{figure}

\begin{figure}
   \begin{center}
       \FigureFile(80mm,0mm){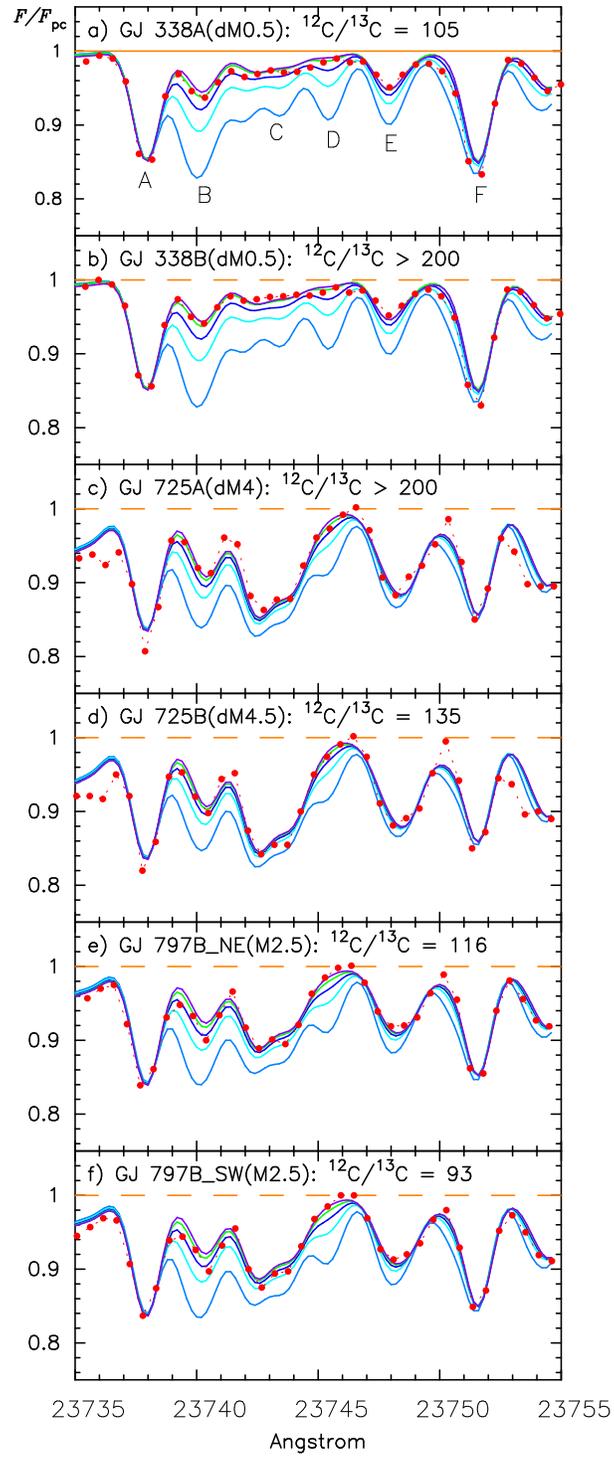}
   \end{center}
   \caption{
    The same as in Fig.\,2, but compare the results for binary pairs:
a) GJ\,338A and b) GJ\,338B, c) GJ\,725A and d) GJ\,725B, and
e) GJ\,797B-NE and f) GJ\,797B-SW. Other details in each panel
are the same as in Fig.\,2.  
}
\label{figure4}
\end{figure}

\begin{figure}
   \begin{center}
       \FigureFile(155mm,0mm){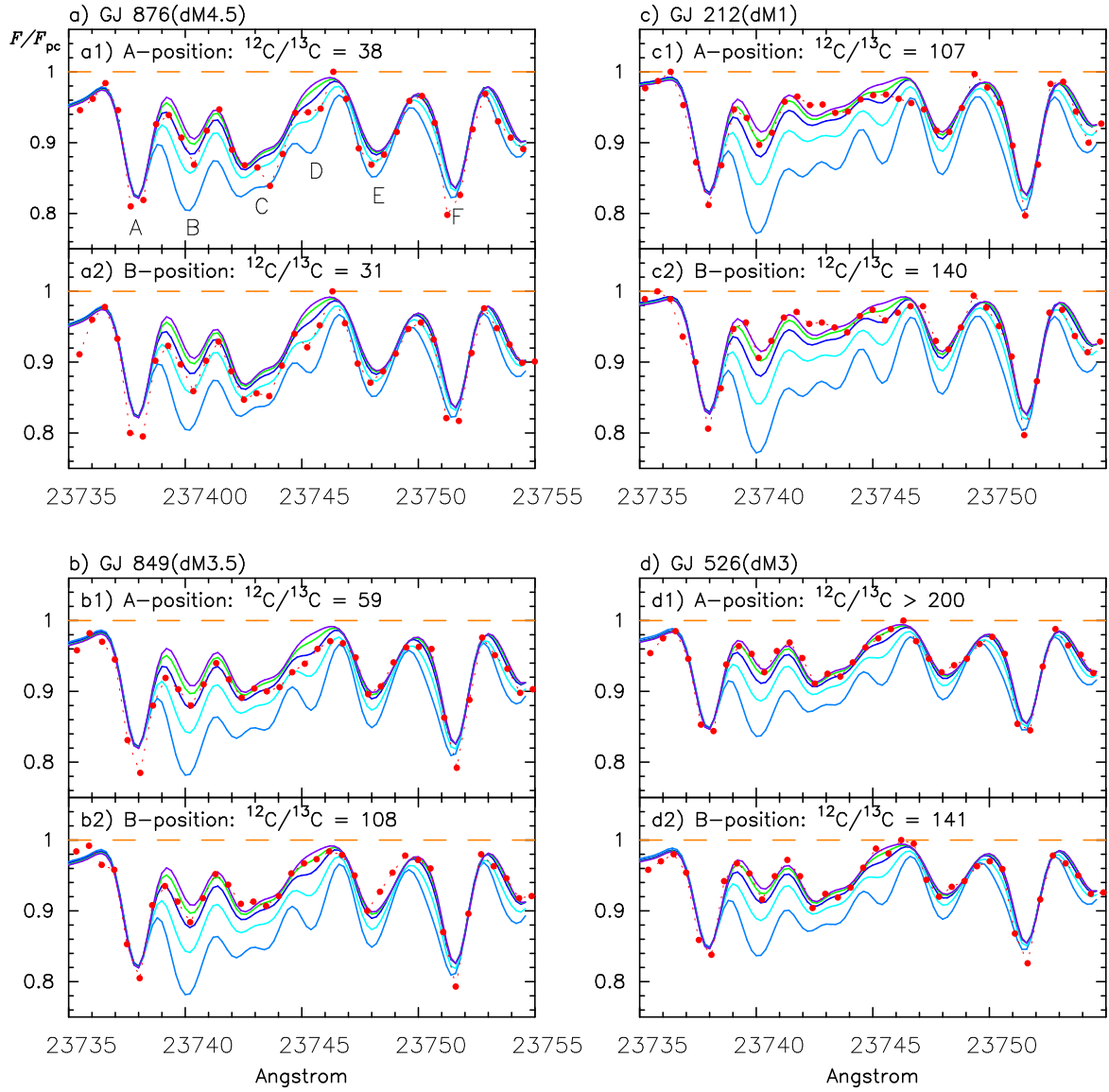}
   \end{center}
   \caption{
    The same as in Fig.\,4, but compare the results for the spectra
observed at positions A and B on the slit:
a) GJ\,876, b) GJ\,849, c) GJ\,212, and d) GJ\,526.
In each object, upper panel shows the result for the spectrum observed
at position A and the lower panel at position B.
}
\label{figure5}
\end{figure}

\begin{figure}
   \begin{center}
       \FigureFile(75mm,0mm){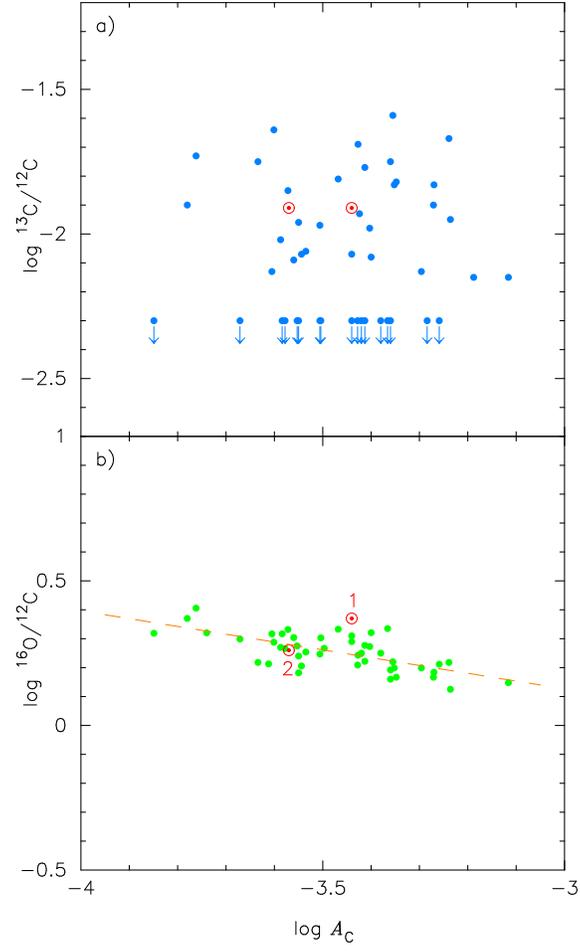}
   \end{center}
   \caption{
   a) The values of  log\,$^{13}$C/$^{12}$C from
Table 3 are plotted against  log\,$A_{\rm C}$. Arrows indicate upper 
limits. The case of the solar photosphere marked by $\odot$ is a mean 
value by \citet{Har87} and by \citet{Ayr06}.  
   b) The values of  log\,$^{16}$O/$^{12}$C (= log\,$A_{\rm O}/A_{\rm C}$ 
from Table 1) are plotted against  log\,$A_{\rm C}$. The dashed line is
a least square fit to the 46 data (equation 1 in subsection 2.2) 
excluding the solar data.
Two cases for the solar photosphere by \citet{And89} and 
by \citet{Asp09} are shown by $\odot$ attached 1 and 2, respectively. 
}
\label{figure6}
\end{figure}

\newpage


\begin{table*}
\begin{center}
\caption{Fundamental parameters, model photospheres, and C \& O abundances
  for 50 M dwarfs. }
\begingroup
\begin{tabular}{llcccccc}
\hline
\noalign{\smallskip}
\noalign{\smallskip}
obj.        & Sp. Type\footnotemark[$*$]  & $ T_{\rm eff}$\footnotemark
[$\dagger$] &
  log\,$g$\footnotemark[$\dagger$] & model\footnotemark[$\ddagger$]
&  log\,$A_{\rm C} \pm {\rm p.e.}$\footnotemark[$\dagger$] &
log\,$A_{\rm O} \pm {\rm p.e.}$\footnotemark[$\S$] &
log\,${A_{\rm O}/A_{\rm C}}$ \\
\noalign{\smallskip}
\hline
\noalign{\smallskip}
GJ\,15A     & dM2.5  &  3567 &   4.890  &  Ca3570c489  &   -3.60 $\pm$ 0.11 &  -3.31 $\pm$ 0.01 &    0.29   \\
GJ\,54.1    & dM5.5e  & 3162   &  5.083 &  Ca3160c508   & -3.38 $\pm$ 0.09  & -3.13 $\pm$ 0.06 & 0.25 \\
GJ\,105B    & dM4.5  &  3360 &   4.954  &  Ca3360c495  &   -3.47 $\pm$ 0.06 &   -3.14 $\pm$ 0.04 &    0.33 \\
GJ\,166C    & dM4e   &  3337 &   4.972  &  Ca3340c497  &   -3.37 $\pm$ 0.14 &   -3.03 $\pm$ 0.06 &    0.34  \\
GJ\,176     & dM2.5e &  3616 &   4.804  &  Ca3620c480  &   -3.35 $\pm$ 0.07 &   -3.15 $\pm$ 0.02 &    0.20  \\
GJ\,179     & dM3.5e &  3476 &   4.877  &  Ca3480c488 &   -3.43 $\pm$ 0.11 &   -3.18 $\pm$ 0.04 &    0.24  \\
GJ\,205     & dM3    &  3801 &   4.710  &  Ca3800c471  &   -3.12 $\pm$ 0.08 &   -2.97 $\pm$ 0.05 &    0.15   \\
GJ\,212     & dM1    &  3757 &   4.748  &  Ca3760c475 &   -3.30 $\pm$ 0.11 &   -3.10 $\pm$ 0.03 &    0.20  \\
GJ\,229     & dM2.5  &  3707 &   4.766  &  Ca3710c477 &   -3.27 $\pm$ 0.07 &   -3.10 $\pm$ 0.02 &    0.17   \\
GJ\,231.1B  & M3.5   &  3442 &   4.897  &  Ca3440c490  &   -3.57 $\pm$ 0.05 &   -3.24 $\pm$ 0.05 &    0.33    \\
\noalign{\smallskip}
GJ\,250B    & M2     &  3567 &   4.827  &  Ca3570c483  &   -3.41 $\pm$ 0.10 &   -3.19 $\pm$ 0.03 &    0.22  \\
GJ\,273     & dM4    &  3415 &   4.915  &  Ca3420c492  &   -3.40 $\pm$ 0.11 &   -3.13 $\pm$ 0.04 &    0.27    \\
GJ\,324B    & M4     &  3382 &   4.938  &  Ca3380c494  &   -3.36 $\pm$ 0.13 &   -3.17 $\pm$ 0.04 &    0.19   \\
GJ\,338A    & dM0.5  &  3907 &   4.709  &  Ca3910c471  &   -3.59 $\pm$ 0.04 &   -3.32 $\pm$ 0.03 &    0.27    \\
GJ\,338B    & dM0.5  &  3867 &   4.709  &  Ca3870c471 &   -3.58 $\pm$ 0.04 &   -3.31 $\pm$ 0.06 &    0.27    \\
GJ\,380     & dM0.5  &  4081 &   4.643  &  Ca4080c464 &   -3.28 $\pm$ 0.06 &   -3.08~~~~~~~~~~  &    0.20     \\
GJ\,406     & dM6.5e &  2800 &   5.170  &  Cc2800c517 &   -3.61 $\pm$ 0.10 &   -3.40 $\pm$ 0.03 &    0.21     \\
GJ\,411     & dM2    &  3465 &   4.857  &  Cc3470c486  &   -3.67 $\pm$ 0.06 &  -3.37 $\pm$ 0.03 &    0.30   \\
GJ\,412A    & dM2    &  3497 &   4.843  &  Cc3500c484 &   -3.85 $\pm$ 0.04 &   -3.53 $\pm$ 0.03 &    0.32   \\
GJ\,436     & dM3.5  &  3416 &   4.797  &  Cc3420c480 &   -3.63 $\pm$ 0.06 &  -3.42 $\pm$ 0.02 &    0.22  \\
\noalign{\smallskip}
GJ\,526     & dM3    &  3618 &   4.784  &  Cc3620c478  &   -3.55 $\pm$ 0.04 &  -3.28 $\pm$ 0.03 &    0.27   \\
GJ\,581     & dM4    &  3442 &   4.959  &  Ca3440c496 &   -3.56 $\pm$ 0.05 &   -3.26 $\pm$ 0.03 &    0.30     \\
GJ\,611B    & M4     &  3202 &   5.063  &  Cc3200c506  &   -3.76 $\pm$ 0.03 &  -3.36 $\pm$ 0.06 &    0.41  \\
GJ\,649     & dM2    &  3660 &   4.784  &  Ca3660c478  &   -3.54 $\pm$ 0.04 &   -3.34 $\pm$ 0.03 &    0.21    \\
GJ\,686     & dM1    &  3538 &   4.842  &  Ca3540c484  &   -3.50 $\pm$ 0.04 &   -3.23 $\pm$ 0.03 &    0.27     \\
GJ\,687     & dM4    &  3413 &   4.811  &  Ca3410c481  &   -3.43 $\pm$ 0.09 &  -3.22 $\pm$ 0.03 &    0.21   \\
GJ\,725A    & dM4    &  3407 &   4.837  &  Cc3410c484  &   -3.58 $\pm$ 0.09 &  -3.27 $\pm$ 0.05 &    0.32    \\
GJ\,725B    & dM4.5  &  3337 &   4.972  &  Cc3340c497  &   -3.61 $\pm$ 0.08 &  -3.29 $\pm$ 0.06 &    0.32    \\
GJ\,752B    & M8     &  2637   &  5.213 &  Cc2640c521  & -3.55 $\pm$ 0.07  & -3.31 $\pm$ 0.04 & 0.24 \\
GJ\,768.1C  &        &  3470 &   4.881  &  Ca3470c488  &   -3.50 $\pm$ 0.08 &  -3.20 $\pm$ 0.04 &    0.30    \\
\noalign{\smallskip}
GJ\,777B    & M4.5   &  3310 &   4.991  &  Ca3310c499   &   -3.24 $\pm$ 0.16 &  -3.02 $\pm$ 0.05 &    0.22   \\
GJ\,783.2B  & M4     &  3370 &   4.949  &  Ca3370c495  &   -3.41 $\pm$ 0.10 &   -3.14 $\pm$ 0.05 &    0.28    \\
GJ\,797B-NE & M2.5   &  3473 &   4.878  &  Ca3470c488  &   -3.54 $\pm$ 0.09 &   -3.28 $\pm$ 0.03 &    0.25  \\
GJ\,797B-SW & M2.5   &  3473 &   4.878  &  Ca3470c488 &   -3.51 $\pm$ 0.09 &   -3.26 $\pm$ 0.03 &    0.25    \\
GJ\,809     & dM2    &  3692 &   4.720  &  Ca3690c472  &   -3.55 $\pm$ 0.04 &   -3.37 $\pm$ 0.03 &    0.18    \\
GJ\,820B    & dM0    &  3932 &   4.679  &  Ca3930c468  &   -3.51 $\pm$ 0.05  &   -3.25~~~~~~~~~~   &    0.26   \\
GJ\,849     & dM3.5  &  3580 &   4.821  &  Ca3580c482  &   -3.27 $\pm$ 0.09 &   -3.09 $\pm$ 0.02 &    0.18   \\
GJ\,873    &  dM4.5e & 3434    &   4.903 &  Ca3430c490 & -3.44 $\pm$ 0.09  & -3.13 $\pm$ 0.06 & 0.31 \\
GJ\,876     & dM4.5  &  3458 &   4.888  &  Ca3460c489  &   -3.36 $\pm$ 0.13 &  -3.14 $\pm$ 0.04 &    0.22     \\
GJ\,880     & dM2.5  &  3713 &   4.716  &  Ca3710c472  &   -3.35 $\pm$ 0.07 &  -3.18 $\pm$ 0.03 &    0.17   \\
\noalign{\smallskip}
GJ\,884     & dM0.5  &  3850 &   4.720  &  Ca3850c472  &   -3.42 $\pm$ 0.07 &   -3.18~~~~~~~~~~   &   0.24   \\
GJ\,1002   & M5.5    & 2974    &  5.123  &  Ca2970c513 & -3.42 $\pm$ 0.09  & -3.17 $\pm$ 0.05 & 0.25 \\
GJ\,1245B  & M5.5    & 2944    &  5.138  & Ca2940c514  & -3.44 $\pm$ 0.10  & -3.15 $\pm$ 0.07 & 0.29 \\
GJ\,3348B   & M4     &  3476 &   4.876  &  Ca3480c488  &   -3.40 $\pm$ 0.11 &  -3.08 $\pm$ 0.05 &    0.32  \\
HIP\,12961  & M0     &  3890 &   4.709  &  Ca3890c471  &   -3.19 $\pm$ 0.07 &  -3.01~~~~~~~~~~  &    0.18     \\
HIP\,57050  & M4     &  3464 &   4.884  &  Ca3460c488  &   -3.26 $\pm$ 0.12 &  -3.05 $\pm$ 0.04 &    0.21   \\
HIP\,79431  & M3     &  3592 &   4.815  &  Ca3590c482 &   -3.24 $\pm$ 0.12 &   -3.11 $\pm$ 0.01 &    0.12 \\
GAT\,1370  & M8/9     & 2685     &  5.199  &  Cc2690c520 & -3.78 $\pm$ 0.06  & -3.41 $\pm$ 0.06 & 0.37 \\
LP\,412-31 & M8     & 2607     &  5.217  & Ca2610c522  & -3.36 $\pm$ 0.07  & -3.20 $\pm$ 0.03 & 0.16 \\
2MASS 1835+32 & M8.5  & 2275   & 5.261   & Bc2280c526  & -3.74 $\pm$ 0.13  & -3.42 $\pm$ 0.02 & 0.32 \\
\noalign{\smallskip}
the Sun 1\footnotemark[$\|$] &      &  　  &     &     & -3.44 $\pm$ 0.04  &  -3.07 $\pm$ 0.035  & 0.37 \\
the Sun 2\footnotemark[$\#$] &      &      &     &     & -3.57 $\pm$ 0.05  & -3.31 $\pm$ 0.05   & 0.26 \\
\noalign{\smallskip}
\hline
   \multicolumn{5}{@{}l@{}}{\hbox to 0pt{\parbox{180mm}{\footnotesize
      \hbox{}
      \par\noindent
      \footnotemark[$*$] Spectral types beginning with dM are from \citet{Joy74} and those beginning with M from SIMBAD.
      \par\noindent
      \footnotemark[$\dagger$]  from Paper I or III.
      \par\noindent
      \footnotemark[$\ddagger$]  Specified model for $T_{\rm eff}$ and
         log\,$g$ in columns 3 and 4, respectively, from Paper I or III.
      \par\noindent
      \footnotemark[$\S$]  from Paper II or III. Those without p.e. are
         estimated from log\,$A_{\rm C}$ based on log\,$A_{\rm O}/A_{\rm C}$ by equation 1.
       \par\noindent
      \footnotemark[$\|$]　\citet{And89}.
      \par\noindent
      \footnotemark[$\#$]  \citet{Asp09}.
      }\hss}}
  \end{tabular}
\endgroup
\end{center}
\end{table*}

\begin{table*}
\begin{center}
\caption{Comparison of the observed and computed line positions of H$_2$O. }
\renewcommand{\arraystretch}{1.28}
\begin{tabular}{rcccccccrrrrrr}
\hline
\noalign{\smallskip}
\noalign{\smallskip}
no. & $ \nu_{\rm obs}$(cm$^{-1}$) \footnotemark[$*$] & 
$ \nu_{\rm cal}$(cm$^{-1}$) \footnotemark[$\dagger$] &
 $ \lambda_{\rm obs}{\rm (\AA)} $\footnotemark[$\ddagger$] & $ \lambda_{\rm cal}{\rm (\AA)} $\footnotemark[$\ddagger$] &
diff. & $v_1^{'}v_2^{'}v_3^{'}$ & $v_1^{''}v_2^{''}v_3^{''}$ &
$J^{'}$ & $K_a^{'}$ & $K_c^{'}$ & $J^{''}$ & $K_a^{''}$ & $K_c^{''}$  \\
\noalign{\smallskip}
\hline
  1 &  4203.8462 & 4203.8450 &  23787.740 &  23787.747 & $-$0.007 &     1\,0\,0 & 0\,0\,0 & 17 & 12 & 5 & 16 & 11 & 6\\  
  2 &  4204.2840 & 4204.2750 &  23785.263 &  23785.314 & $-$0.051 &     1\,0\,0 & 0\,0\,0 & 17 & 16 & 1 & 16 & 15 & 2\\  
  3 &  4205.4124 & 4205.4110 &  23778.881 &  23778.889 & $-$0.008 &     1\,0\,0 & 0\,0\,0 & 17 &  9 & 9 & 16 & 8  & 8\\  
  4 &  4206.7835 & 4206.7830 &  23771.131 &  23771.133 & $-$0.003 &     1\,0\,0 & 0\,0\,0 & 17 &  9 & 8 & 16 & 8  & 9\\  
  5 &  4208.4149 & 4208.4120 &  23761.916 &  23761.932 & $-$0.016 &     1\,0\,0 & 0\,0\,0 & 17 & 14 & 3 & 16 & 13 & 4\\  
  6 &  4211.5887 & 4211.5830 &  23744.009 &  23744.041 & $-$0.032 &     1\,0\,0 & 0\,0\,0 & 19 & 10 & 9 & 18 & 9  & 10\\  
  7 &  4211.7144 & 4211.7110 &  23743.300 &  23743.320 & $-$0.019 &     1\,0\,0 & 0\,0\,0 & 18 & 18 & 1 & 17 & 17 & 0\\  
  8 &  4211.8684 & 4211.8610 &  23742.432 &  23742.474 & $-$0.042 &     1\,0\,0 & 0\,0\,0 & 18 & 11 & 8 & 17 & 10 & 7\\  
  9 &  4214.3640 & 4214.3610 &  23728.373 &  23728.390 & $-$0.017 &     1\,0\,0 & 0\,0\,0 & 15 &  5 & 10 & 14 & 4 & 11\\  
 10 &  4220.0551 & 4220.0530 &  23696.373 &  23696.385 & $-$0.012 &     1\,0\,0 & 0\,0\,0 & 18 & 9 & 10  & 17 & 8 & 9\\  
 11 &  4220.5941 & 4220.5860 &  23693.347 &  23693.392 & $-$0.045 &     1\,0\,0 & 0\,0\,0 & 18 & 16 & 3 & 17 & 15 & 2\\  
 12 &  4226.5447 & 4226.5450 &  23659.989 &  23659.987 & $+$0.002 &     1\,0\,0 & 0\,0\,0 & 19 & 18 & 1 & 18 & 17 & 2\\  
 13 &  4234.2558 & 4234.2450 &  23616.901 &  23616.961 & $-$0.060 &     1\,0\,0 & 0\,0\,0 & 19 & 12 & 7 & 18 & 11 & 8\\  
 14 &  4238.7311 & 4238.7270 &  23591.966 &  23591.989 & $-$0.023 &     1\,0\,0 & 0\,0\,0 & 19 & 15 & 4 & 18 & 14 & 5\\  
\noalign{\smallskip}
\hline
   \multicolumn{7}{@{}l@{}}{\hbox to 0pt{\parbox{180mm}{\footnotesize
      \hbox{}
      \par\noindent
      \footnotemark[$*$] measured from the digital version of the Sunspot Umbral Spectrum \citep{Wal92}.
      \par\noindent
      \footnotemark[$\dagger$] from BT2-HITEMP2010 (\cite{Bar06}; \cite{Rot10}).
      \par\noindent
      \footnotemark[$\ddagger$] in vacuum.
      }\hss}}
  \end{tabular}
\end{center}
\end{table*}

\begin{table*}
\begin{center}
\caption{$\chi^2_r(^{12}{\rm C}/^{13}{\rm C})$ for five values of $^{12}{\rm C}/^{13}{\rm C}$,
 $^{12}{\rm C}/^{13}{\rm C}$ for the $\chi^2_r$ minimum, log\,$^{13}{\rm C}/^{12}{\rm C}$
 and S/N ratio.
 }
\begingroup
\begin{tabular}{lrrrrrccr}
\hline
\noalign{\smallskip}
\noalign{\smallskip}
obj.        & $\chi^2_r(10)$  & $\chi^2_r(25)$  & $\chi^2_r(50) $ &$\chi^2_r(100)$ & $\chi^2_r(200) $
 & $^{12}{\rm C}/^{13}{\rm C}(\chi^2_{\rm min})$\footnotemark[$*$] & 
log\,$^{13}{\rm C}/^{12}{\rm C}$ &  S/N \\
\noalign{\smallskip}
\hline
\noalign{\smallskip}
GJ\,15A  &        7.913 &    1.851  &   1.648 &    2.108 &    2.515 &   ~~~44  &  $~~~-$1.64 & 78\\
GJ\,54.1 &       22.076 &    7.466 &    4.739 &    3.918 &    3.698 &   $>$200 &  $<-$2.30   & 45\\
GJ\,105B &        9.951 &    4.088 &    3.416 &    3.536 &    3.775 &    ~~~64 &  $~~~-$1.81 & 70 \\
GJ\,166C &       18.002 &    7.224 &    5.155 &    4.512 &    4.345 &   $>$200 &  $<-$2.30   & 63\\
GJ\,176  &       23.169 &    4.497 &    1.891 &    2.078 &    2.744 &    ~~~68 &  $~~~-$1.83 & 71\\
GJ\,179  &       13.845 &    3.605 &    3.072 &    3.705 &    4.361 &    ~~~49 &  $~~~-$1.69 & 68 \\
GJ\,205  &       67.790 &   21.498 &    7.137 &    3.194 &    3.241 &    ~~140 &  $~~~-$2.15 & 64\\
GJ\,212  &       41.244 &   10.138 &    2.674 &    1.505 &    1.607 &    ~~134 &  $~~~-$2.13 & 63\\
GJ\,229  &       35.496 &    7.866 &    2.706 &    2.525 &    3.363 &    ~~~80 &  $~~~-$1.90 & 64\\
GJ\,231.1B &     10.690 &    2.546 &    1.638 &    1.644 &    1.808 &    ~~~70 &  $~~~-$1.85 & 126\\
\noalign{\smallskip}
GJ\,250B     &   28.015  &   9.063  &   3.434  &   2.327  &   2.153  &  $>$200 &  $<-$2.30   & 72\\
GJ\,273     &   15.595   &  4.217   &   2.657  &   2.517  &   2.701  &   ~~~95 &  $~~~-$1.98 & 69\\
GJ\,324B    &    15.074  &   3.276  &   2.168  &   2.730  &   3.400  &   ~~~56 &  $~~~-$1.75 & 120\\
GJ\,338A    &    21.385  &   3.691  &   0.964  &   0.922  &   0.954  &   ~~105 &  $~~~-$2.02 & 114\\
GJ\,338B    &    24.078  &   4.740  &   1.249  &   0.751  &   0.690  &  $>$200 &  $<-$2.30   & 120\\
GJ\,380     &    76.979  &  24.081  &   8.305  &   3.181  &   1.908  &  $>$200 &  $<-$2.30   &  59\\
GJ\,406     &    25.263  &   6.778  &   3.754  &   3.600  &   3.850  &   ~~~92 &  $~~~-$1.96 &  55\\
GJ\,411     &    11.006  &   2.372  &   1.123  &   0.875  &   0.874  &  $>$200 &  $<-$2.30   & 119\\
GJ\,412A    &     8.083  &   3.692  &   3.053  &   2.910  &   2.894  &  $>$200 &  $<-$2.30   & 135\\
GJ\,436     &     9.841  &   2.158  &   1.537  &  1.852   &  2.215   &   ~~~56 &  $~~~-$1.95 & 109\\
\noalign{\smallskip}
GJ\,526     &    21.705  &   6.160 &    1.905  &   0.820  &   0.510  &  $>$200 &  $<-$2.30   & 118\\
GJ\,581     &    12.142  &   3.158 &    1.878  &   1.672  &   1.721  &   ~~124 &  $~~~-$2.09 & 105\\
GJ\,611B    &     5.685  &   3.593 &    3.427  &   3.534  &   3.649  &   ~~~54 &  $~~~-$1.73 &  95\\
GJ\,649     &    20.342  &   3.983 &    1.386  &   1.069  &   1.188  &   ~~117 &  $~~~-$2.07 & 114\\
GJ\,687     &    28.257  &   8.680 &    3.838  &   2.771  &   2.570  &  $>$200 &  $<-$2.30   & 124\\
GJ\,725A    &    14.798  &   5.524 &    3.693  &   3.163  &   3.041  &  $>$200 &  $<-$2.30   & 165\\
GJ\,725B    &    10.967  &   4.702 &    3.657  &   3.453  &  3.467   &   ~~135 &  $~~~-$2.13 &  91\\
GJ\,752B    &    38.982  &  13.280 &    7.916  &   6.825  &   6.408  &  $>$200 &  $<-$2.30   &  64\\
GJ\,768.1C  &    21.125  &   5.977 &    3.398  &   2.672  &   2.493  &  $>$200 &  $<-$2.30　 &　86\\
GJ\,777B    &    14.325  &   6.440 &    6.015  &   6.619  &   7.302  &   ~~~47 &  $~~~-$1.67 &  73\\
\noalign{\smallskip}
GJ\,783.2B  &    11.072  &   3.796 &    3.128  &   3.355  &   3.705  &   ~~~59 &  $~~~-$1.77 &  152\\
GJ\,797B-NE &    13.402  &   2.828 &    1.129  &   0.892  &   0.989  &   ~~116 &  $~~~-$2.06 &  130\\
GJ\,797B-SW &    12.733  &   2.467 &    0.936  &   0.821  &   0.998  &   ~~~93 &  $~~~-$1.97 &  130\\
GJ\,809     &    23.483  &   4.948 &    1.809  &   1.659  &   1.898  &   ~~~92 &  $~~~-$1.96 &  94\\
GJ\,820B    &    38.247  &   9.876 &    3.407  &   1.700  &   1.375  &  $>$200 &  $<-$2.30   &  77\\
GJ\,849     &    23.646  &   4.981 &    1.376  &   1.725  &   2.617  &   ~~~67 &  $~~~-$1.83 &  72\\
GJ\,873     &    15.949  &   4.170 &    2.456  &   2.197  &   2.292  &   ~~117 &  $~~~-$2.07 &  46\\
GJ\,876     &    13.128  &   2.979 &    2.833  &   3.843  &   4.775  &   ~~~39 &  $~~~-$1.59 &  71\\
GJ\,880     &    30.162  &   5.856 &    2.273  &   2.655  &  3.593   &   ~~~66 &  $~~~-$1.82 &  70\\
GJ\,884     &    29.031  &   6.078 &    2.398  &   2.229  &   2.686  &   ~~~85 &  $~~~-$1.93 &  62\\
\noalign{\smallskip}
GJ\,1002    &    22.928  &   6.077 &    3.299  &   2.642  &   2.560  &  $>$200 &  $<-$2.30   &  76\\
GJ\,1245B   &    24.179  &   8.541 &    6.013  &   5.440  &   5.366  &  $>$200 &  $<-$2.30   &  45\\
GJ\,3348B   &    16.424  &   3.878 &    1.980  &   1.676  &   1.775  &   ~~119 &  $~~~-$2.08 &  99\\
HIP\,12961  &    60.679  &  17.211 &    4.949  &   2.044  &   2.092  &   ~~140 &  $~~~-$2.15 &  59\\
HIP\,57050  &    34.777  &  10.965 &    4.118  &   2.202  &   1.625  &  $>$200 &  $<-$2.30   & 140\\
HIP\,79431  &    34.252  &   7.670 &    2.597  &   2.195  &   2.957  &   ~~~90 &  $~~~-$1.95 &  93\\
GAT\,1370   &    14.485  &   9.067 &    8.327  &   8.178  &   8.938  &   ~~~79 &  $~~~-$1.90 &  69\\
LP\,412-31  &    71.340  &  23.198 &    9.200  &   4.081  &   3.780  &  $>$200 &  $<-$2.30   &  58\\
the Sun\footnotemark[$\dagger$]     &      &      &     &     &    &   ~~~82  & $~~~-$1.91  &    \\
\noalign{\smallskip}
\hline
   \multicolumn{5}{@{}l@{}}{\hbox to 0pt{\parbox{180mm}{\footnotesize
      \hbox{}
      \par\noindent
      \footnotemark[$*$]  $^{12}{\rm C}/^{13}{\rm C}$  at which $\chi^2_r$ is minimum.
      \par\noindent
      \footnotemark[$\dagger$]  $ (^{12}{\rm C}/^{13}{\rm C})_{\rm Sun}$
         is the mean value of
         the solar photosphere by \citet{Har87} and by \citet{Ayr06}.
      }\hss}}
  \end{tabular}
\endgroup
\end{center}
\end{table*}

\begin{table}
\begin{center}
\caption{$\chi^2_r(^{12}{\rm C}/^{13}{\rm C})$ for five values of $^{12}{\rm C}/^{13}{\rm C}$,
and $^{12}{\rm C}/^{13}{\rm C}$ for the $\chi^2_r$ minimum.
 }
\begingroup
\renewcommand{\arraystretch}{1.12}
\begin{tabular}{lrrrrrc}
\hline
\noalign{\smallskip}
\noalign{\smallskip}
obj.\footnotemark[$*$] & $\chi^2_r(10)$  & $\chi^2_r(25)$  & $\chi^2_r(50) $ &$\chi^2_r(100)$ 
& $\chi^2_r(200)$ & $^{12}{\rm C}/^{13}{\rm C}(\chi^2_{\rm min})$\footnotemark[$\dagger$] \\
\noalign{\smallskip}
\hline
\noalign{\smallskip}                                                          
GJ\,876-A &   14.009  &   3.840  &   3.746 &    4.726 &    5.638  & ~~~37.7 \\
GJ\,876-B &   11.143  &   3.385  &   4.437 &    5.962 &    7.173  & ~~~31.4 \\
\noalign{\smallskip}
GJ\,849-A &   21.499  &   4.654  &   2.335 &    3.140 &    4.282  & ~~~59.3 \\
GJ\,849-B &   30.090  &   8.070  &   2.104 &    1.516 &    1.883  & ~~108.3  \\
\noalign{\smallskip}
GJ\,212-A &   36.859  &   8.519  &   2.591 &    1.915 &    2.371  & ~~106.9 \\
GJ\,212-B &   42.102  &  11.035  &   3.537 &    2.211 &    2.232  & ~~139.7 \\
\noalign{\smallskip}
GJ\,526-A &   18.447  &   4.640  &   1.514 &    0.901 &    0.836  &  $>$200.0 \\
GJ\,526-B &   17.125  &   4.459  &   1.812 &    1.351 &    1.352  &  ~~141.0 \\
\noalign{\smallskip}
\hline
   \multicolumn{5}{@{}l@{}}{\hbox to 0pt{\parbox{90mm}{\footnotesize
      \hbox{}
      \par\noindent
      \footnotemark[$*$] A and B after object name refer to the positons A and B on the slit
         and should not be confused with the binary componets. 
      \par\noindent
      \footnotemark[$\dagger$] $^{12}{\rm C}/^{13}{\rm C}$  at which $\chi^2_r$ is minimum.  
      }\hss}}
  \end{tabular}
\endgroup
\end{center}
\end{table}

\begin{table*}
\begin{center}
\caption{ $^{12}{\rm C}/^{13}{\rm C}$ ratios based on separated A, B, and 
co-added spectra. }
\renewcommand{\arraystretch}{1.28}
\begin{tabular}{lcccc}
\hline
\noalign{\smallskip}
\noalign{\smallskip}
obj. &  $(^{12}{\rm C}/^{13}{\rm C})_{\rm A}$\footnotemark[$*$] &
 $(^{12}{\rm C}/^{13}{\rm C})_{\rm B}$\footnotemark[$\dagger$] &   
  $(^{12}{\rm C}/^{13}{\rm C})_{\rm mean} \pm {\rm p.e.}$
\footnotemark[$\ddagger$] &
  $(^{12}{\rm C}/^{13}{\rm C})_{\rm A+B}$\footnotemark[$\S$] \\
\noalign{\smallskip}
\hline
\noalign{\smallskip}
GJ876  &    ~~~37.7 & ~31.4 &  ~~~34.6 $\pm$ ~3.0  &   ~~~39 \\

GJ849  &    ~~~59.3 & 108.3 &  ~~~83.8 $\pm$ 23.4  &     ~~~67\\

GJ212  &     ~~106.9 & 139.7  & ~~123.3 $\pm$ 15.6   &  ~~134 \\

GJ526  &  $>$200.0 & 141.0  &    & $>$200 \\
\noalign{\smallskip}
\hline
   \multicolumn{5}{@{}l@{}}{\hbox to 0pt{\parbox{150mm}{\footnotesize
      \hbox{}
      \par\noindent
      \footnotemark[$*$]  based on the spectrum at position A (from Table 4).
      \par\noindent
      \footnotemark[$\dagger$] based on the spectrum at position B (from 
           Table 4).
      \par\noindent
      \footnotemark[$\ddagger$] mean of the results based on the spectra
         at positions A and B.
      \par\noindent
      \footnotemark[$\S$]  based on the coadded spectrum (from Table 3).
      }\hss}}
  \end{tabular}
  \end{center}
\end{table*}

\end{document}